\documentclass[12pt,a4paper]{article}

\pdfoutput=1
\usepackage{jheppub,psfrag,slashed,cancel,lscape,caption,array,graphicx,subcaption}
\usepackage[utf8]{inputenc}
\usepackage{amsmath}
\usepackage{nicefrac}  
\usepackage{mathtools}
\usepackage{mathrsfs}
\usepackage{multirow}
\usepackage{booktabs} 
\usepackage{chngcntr}
\usepackage{nicefrac}
\usepackage{cleveref}
\usepackage{orcidlink}
\usepackage{relsize}
\usepackage[normalem]{ulem}

\usepackage{physics}

\usepackage{url}
\usepackage{hyperref}
\usepackage{tikz}
\usetikzlibrary{decorations.pathmorphing}
\usetikzlibrary{decorations.markings}
\usetikzlibrary{positioning, shapes, snakes, arrows}
\usetikzlibrary{patterns.meta}
\usetikzlibrary{arrows.meta}
\usetikzlibrary{calc}
\usetikzlibrary{backgrounds}

\pgfdeclarelayer{foreground}
\pgfsetlayers{background,main,foreground}

\tikzset{
  externalShorter/.style={
    shorten <=2mm,
    shorten >=2mm
  },
  wl/.style={line width=1pt},
  graviton/.style={line width=.8pt, -latex,decorate, decoration={snake, segment length=4pt,amplitude=1.8pt, pre length=.15cm, post length=.25cm}},
    worldlineStatic/.style={dotted, line width=1pt},
	worldline/.style={gray, line width=1pt},
	worldlineBold/.style={black, line width=.6pt},
	zUndirected/.style={line width=1pt},
	zParticle/.style={line width=1pt,postaction={decorate},decoration={markings,mark=at position .6 with {\arrow[#1]{latex}}}},
	zParticleTest/.style={white,line width=1pt,postaction={decorate},decoration={markings,mark=at position .6 with {\arrow[#1]{latex}}}},
  example1/.style={
    draw=none,
    postaction={decorate},
    decoration={
      markings,
      mark=at position #1 with {
        \arrow[
          black,
          scale=2.4
        ]{Straight Barb}
      }
    }
  },
  example2/.style={
    white,
    line width=1pt,
    postaction={decorate},
    decoration={
      markings,
      mark=at position .6 with {
        \arrow[
          draw=black,
          scale=1.5,
          #1
          ]{latex}
      }
    }
  },
	zParticle2/.style={line width=1pt,postaction={decorate},decoration={markings,mark=at position .75 with {\arrow[#1]{latex}}}},
	zParticle1/.style={line width=1pt},
	worldlineCut/.style={dotted,line width=1pt,postaction={decorate},decoration={markings,mark=at position .7 with {\arrow[#1]{latex}}}},
	worldlineCut2/.style={dotted,line width=1pt,postaction={decorate},decoration={markings,mark=at position .6 with {\arrow[#1]{latex}}}},
	zParticleF/.style={line width=1pt,postaction={decorate}},
	cscalar/.style={line width=1pt,postaction={decorate},decoration={markings,mark=at position .6 with {\arrow[#1]{latex}}}},
	cscalar2/.style={line width=1pt,postaction={decorate},decoration={markings,mark=at position .8 with {\arrow[#1]{latex}}}},
	photon/.style={line width =.8pt, decorate, decoration={snake, segment length=4pt, amplitude=1.8pt,  pre length=.1cm, post length=.1cm}},
	photonTest/.style={
    line width =.8pt,
    decorate,
    decoration={
      snake,
      segment length=5pt,
      amplitude=1.4pt,
      pre length=.0cm,
      post length=.0cm}},
	photonRed/.style={red, line width =.8pt, decorate, decoration={snake, segment length=4pt, amplitude=1.8pt,  pre length=.1cm, post length=.1cm}},
	cross/.style={cross out, line width =.8pt, draw=black, minimum size=2*(#1-\pgflinewidth), inner sep=0pt, outer sep=0pt},
cross/.default={4pt},
  dottedLine/.style={
    dotted, thick
  }
}

\newcommand{\iO}{\i 0^{+}}
\newcommand{\epsDR}{\varepsilon}
\newcommand{\epsPM}{\epsilon_{\rm PM}}

\newcommand{\gravitonLine}[3]{
    \draw [photonTest] #1 -- #2 ;
    \draw [example1={#3}] #1 -- #2 ;
}
\newcommand{\zLine}[2]{
    \draw [zParticle] #1 -- #2 ;
}
\newcommand{\vertexSmall}[1]{
  \begin{pgfonlayer}{foreground}
    \draw[fill] #1 circle (.06);
    \end{pgfonlayer}
    }

\newcommand{\snakeblob}[2]{
  \draw [fill=white,draw=white] #1 circle (0.3cm) ;
  \begin{scope}[shift={#1}, rotate=#2]
    \clip (0,0) circle (0.3cm);
    \foreach \y in {-0.6,-0.5,...,0.6} {
      \draw[
        decorate,
        decoration={snake, amplitude=1pt, segment length=6pt}
      ] (-1,\y) -- (1,\y);
    }
  \end{scope}
}

\newcommand{\drawCompton}[1]{
  \draw [fill=black!60, thick] #1 circle (0.3cm);
}
\newcommand{\drawM}{
  \begin{scope}[shift={(currentLocation)}]
      \draw[photonTest,red] (0,0) to[out=90,in=180] (.4,.4) to[out=0,in=90] (.8,0) ;
      \draw[dottedLine] (0,0) -- (.8,0) ;
      \vertexSmall{(0,0)}
      \vertexSmall{(.8,0)}
  \end{scope}
  \coordinate (currentLocation) at ($(currentLocation)+(.8,0)$) ;
}
\newcommand{\drawWL}{
  \begin{scope}[shift={(currentLocation)}]
    \draw[zParticle1] (0,0) -- (.5,0) ;
    \vertexSmall{(0,0)}
    \vertexSmall{(.5,0)}
  \end{scope}
  \coordinate (currentLocation) at ($(currentLocation)+(.5,0)$) ;
}
\newcommand{\drawMtoLshort}{
  \begin{scope}[shift={(currentLocation)}]
    \draw[photonTest,red] (.6,.8) to[out=180,in=0] (0,0) ;
    \draw[dottedLine] (0,0) -- (.6,0) ;
  \end{scope}
  \coordinate (currentLocation) at ($(currentLocation)+(.6,0)$) ;
}
\newcommand{\drawLtoMshort}{
  \begin{scope}[shift={(currentLocation)}]
    \draw[photonTest,red] (0,.8) to[out=0,in=180] (.6,0) ;
    \draw[dottedLine] (0,0) -- (.6,0) ;
  \end{scope}
  \coordinate (currentLocation) at ($(currentLocation)+(.6,0)$) ;
}
\newcommand{\drawMtoLmedium}{
  \begin{scope}[shift={(currentLocation)}]
    \draw[photonTest,red] (.75,.8) to[out=180,in=0] (0,0) ;
    \draw[dottedLine] (0,0) -- (.75,0) ;
  \end{scope}
  \coordinate (currentLocation) at ($(currentLocation)+(.75,0)$) ;
}
\newcommand{\drawLtoMmedium}{
  \begin{scope}[shift={(currentLocation)}]
    \draw[photonTest,red] (0,.8) to[out=0,in=180] (.75,0) ;
    \draw[dottedLine] (0,0) -- (.75,0) ;
  \end{scope}
  \coordinate (currentLocation) at ($(currentLocation)+(.75,0)$) ;
}
\newcommand{\drawLtoLshort}{
  \begin{scope}[shift={(currentLocation)}]
    \draw[photonTest,red] (0,.8) -- (.6,.8) ;
    \draw[dottedLine] (0,0) -- (.6,0) ;
  \end{scope}
  \coordinate (currentLocation) at ($(currentLocation)+(.6,0)$) ;
}
\newcommand{\drawLzero}{
  \begin{scope}[shift={(currentLocation)}]
    \draw[photonTest] (0,0) -- (0,.8) ;
    \vertexSmall{(0,0)}
    \vertexSmall{(0,.8)}
  \end{scope}
}
\newcommand{\drawLoneA}{
  \begin{scope}[shift={(currentLocation)}]
    \draw[photonTest] (.3,0) -- (0,.8) ;
    \draw[photonTest] (-.3,0) -- (0,.8) ;
    \vertexSmall{(.3,0)}
    \vertexSmall{(-.3,0)}
    \vertexSmall{(0,.8)}
  \end{scope}
}
\newcommand{\drawLoneB}{
  \begin{scope}[shift={(currentLocation)}]
    \draw[photonTest] (.3,0) -- (0,.5) ;
    \draw[photonTest] (-.3,0) -- (0,.5) ;
    \draw[photonTest] (0,.5) -- (0,.8) ;
    \vertexSmall{(.3,0)}
    \vertexSmall{(-.3,0)}
    \vertexSmall{(0,.8)}
    \vertexSmall{(0,.5)}
  \end{scope}
}
\newcommand{\drawLtwoA}{
  \begin{scope}[shift={(currentLocation)}]
    \draw[photonTest] (.4,0) -- (0,.8) ;
    \draw[photonTest] (-.4,0) -- (0,.8) ;
    \draw[photonTest] (0,0) -- (0,.8) ;
    \vertexSmall{(.4,0)}
    \vertexSmall{(0,0)}
    \vertexSmall{(-.4,0)}
    \vertexSmall{(0,.8)}
  \end{scope}
}
\newcommand{\drawLtwoB}{
  \begin{scope}[shift={(currentLocation)}]
    \draw[photonTest] (-.4,0) -- (-.15,.4) ;
    \draw[photonTest] (.1,0) -- (-.15,.4) ;
    \draw[photonTest] (-.15,.4) -- (0,.8) ;
    \draw[photonTest] (.4,0) -- (0,.8) ;
    \vertexSmall{(.4,0)}
    \vertexSmall{(-.15,.4)}
    \vertexSmall{(.1,0)}
    \vertexSmall{(-.4,0)}
    \vertexSmall{(0,.8)}
  \end{scope}
}
\newcommand{\drawLtwoC}{
  \begin{scope}[shift={(currentLocation)}]
    \draw[photonTest] (.4,0) -- (0,.5) ;
    \draw[photonTest] (-.4,0) -- (0,.5) ;
    \draw[photonTest] (0,0) -- (0,.5) ;
    \draw[photonTest] (0,.5) -- (0,.8) ;
    \vertexSmall{(.4,0)}
    \vertexSmall{(0,.8)}
    \vertexSmall{(0,0)}
    \vertexSmall{(-.4,0)}
    \vertexSmall{(0,.5)}
  \end{scope}
}
\newcommand{\drawLtwoD}{
  \coordinate (newLocation) at ($(currentLocation)-(.3,0)$) ;
  \begin{scope}[shift={(newLocation)}]
    \draw[photonTest] (0,0) -- (.15,.3) ;
    \draw[photonTest] (.3,0) -- (.15,.3) ;
    \draw[photonTest] (.15,.3) -- (.3,.6) ;
    \draw[photonTest] (.6,0) -- (.3,.6) ;
    \draw[photonTest] (.3,.6) -- (.5,.8) ;
    \vertexSmall{(0,0)}
    \vertexSmall{(.3,0)}
    \vertexSmall{(.6,0)}
    \vertexSmall{(.15,.3)}
    \vertexSmall{(.3,.6)}
    \vertexSmall{(.5,.8)}
  \end{scope}
}

\DeclareFontFamily{OT1}{pzc}{} 
\DeclareFontShape{OT1}{pzc}{m}{it}{<-> s * [1.350] pzcmi7t}{}
\DeclareMathAlphabet{\mathpzc}{OT1}{pzc}{m}{it}

\def\cO{\mathcal{O}}

\def\cM{\mathcal{M}}

\def\eps{\epsilon}

\def\d{\mathrm{d}}

\def\pat{\partial}
\def\mn{{\mu\nu}}
\def\ab{{\alpha\beta}}

\def\i\math

\def\bH{\hat{b}}

\def\dd{\delta\!\!\!{}^-\!}

\def\d{\mathrm{d}}
\def\eps{\epsilon}

\renewcommand{\i}{\ensuremath{i}}
\renewcommand{\d}{\ensuremath{\mathrm{d}}}

\newcommand{\s}[1]{\relax}

\def\nn{\nonumber}

\newcommand*{\vct}[1]{\boldsymbol{#1}}

\newcommand{\be}{\begin{equation}}
\newcommand{\ee}{\end{equation}}
\newcommand{\ba}{\begin{align}}
\newcommand{\ea}{\end{align}}
\ifx\genfrac\sdflkaj\else\fi

\def\centerarc[#1](#2)(#3:#4:#5){ \draw[#1] ($(#2)+({#5*cos(#3)},{#5*sin(#3)})$) arc (#3:#4:#5); }

\begin{document}

\begin{flushright}
\begingroup\footnotesize\ttfamily
	HU-EP-26/05
\endgroup
\end{flushright}

\vspace{15mm}

\begin{center}

{\LARGE\bfseries 
	Gravitational Wave Scattering in Spinless WQFT
\par}

\vspace{15mm}

\begingroup\scshape\large 
	Yilber Fabian Bautista\,\orcidlink{0000-0001-6255-5675},${}^{1,2}$
    Mathias Driesse\,\orcidlink{0000-0002-3983-5852},${}^{3}$ 
    Kays Haddad\,\orcidlink{0000-0002-1182-2750},${}^{3}$ 
	and Gustav Uhre Jakobsen\,\orcidlink{0000-0001-9743-0442},${}^{3,4}$
\endgroup
\vspace{3mm}

\textit{${}^{1}$Higgs Centre for Theoretical Physics, School of Physics and Astronomy, The University of Edinburgh, Edinburgh EH9 3JZ, Scotland, UK} \\[0.25cm]
\textit{${}^{2}$Institut de Physique Théorique, CEA, Université Paris–Saclay,
F–91191 Gif-sur-Yvette cedex, France} \\[0.25cm]
\textit{${}^{3}$Institut f\"ur Physik, Humboldt-Universit\"at zu Berlin, 
 10099 Berlin, Germany} \\[0.25cm]
\textit{${}^{4}$Max-Planck-Institut f\"ur Gravitationsphysik
(Albert-Einstein-Institut), \\ 14476 Potsdam, Germany }

\bigskip
  
\texttt{\small\{yfabian.bautista@ed.ac.uk, mathias.driesse@physik.hu-berlin.de, kays.haddad@physik.hu-berlin.de, gustav.uhre.jakobsen@physik.hu-berlin.de\}}

\vspace{10mm}

\textbf{Abstract}\vspace{5mm}\par
\begin{minipage}{14.7cm}
  We develop the computational framework for gravitational wave--black hole scattering in worldline quantum field theory (WQFT) without spin.
  Crucially, we prove on general grounds that, in the absence of dissipation, the exponential representation of the $S$-matrix maps -- through a partial-wave transformation -- directly onto the scattering phase shift from black hole perturbation theory (BHPT), indicating an exponentiation of the WQFT amplitude itself in partial-wave space.
  Computing explicitly, we reproduce the BHPT phase shift without spin up to $\cO(G^{3})$ from WQFT.
  While this result is expected, it lays the groundwork for higher-precision analyses involving non-minimal effects.
  Along the way, we outline our efficient diagram generation technique and include a pedagogical discussion on the computation of the required two-loop integrals.
\end{minipage}\par

\end{center}
\setcounter{page}{0}
\thispagestyle{empty}
\newpage

\tableofcontents

\section{Introduction}

The scattering of waves off of black holes is a subject with a long history (see, e.g.,~\cite{Futterman_Handler_Matzner_1988}).
Originally introduced to study wave propagation in curved spacetimes, this topic finds several phenomenological applications.
In strong-field backgrounds, for example, wave propagation is important for understanding the electromagnetic properties of black hole accretion disks.
Such studies formed the foundations of realistic models of matter around black holes, ultimately leading to the first direct image of a black hole by the Event Horizon Telescope~\cite{EventHorizonTelescope:2019dse}.

At the other end of the spectrum, gravitational-wave (GW) studies focus on low-frequency wave phenomena influenced by black holes, which are amenable to perturbation theory.
Indeed, the discovery of GWs from compact-binary mergers by the LIGO/Virgo collaboration~\cite{LIGOScientific:2016aoc} has rejuvenated interest in the weak-field scattering of waves off of black holes~\cite{Bautista:2021wfy,Bautista:2022wjf,Bautista:2023sdf,Bautista:2024emt,Ivanov:2024sds,Correia:2024jgr,Caron-Huot:2025tlq,Correia:2025enx,Combaluzier--Szteinsznaider:2025eoc,Kosmopoulos:2025rfj,Parra-Martinez:2025bcu,Akhtar:2025nmt}.
A robust understanding of this process is increasingly important as analytic control over post-Minkowskian (PM) binary scattering nears sensitivity to gravitational tidal effects \cite{Driesse:2024xad,Driesse:2024feo,Bern:2025zno,Bern:2025wyd,Driesse:2026qiz}.

Within the PM framework, effective field theoretic approaches are often employed in which the black hole is modeled as a massive point particle with the gravitational field mediating interactions (with another compact object or with a massless wave).
In this leading-order approximation, black holes are indistinguishable from other compact objects, such as neutron stars.
Differences emerge when analyzing the tidal response of these bodies to external gravitational fields, an effect which becomes relevant either when the wavelength of the perturbing wave becomes comparable to the object’s size, or at a high enough PM precision.
The body-specific behavior is encoded in the coefficients of multipolar deformations of the theory, the so-called tidal Love numbers~\cite{Goldberger:2006bd,Goldberger:2007hy,Goldberger:2022ebt,Bini_2020,Bern:2020uwk}, which in the static limit are expected to vanish for black holes in four dimensions \cite{Binnington:2009bb,Damour:2009vw,Kol:2011vg,Porto:2016zng,Charalambous:2021kcz,Hui:2021vcv,Charalambous:2022rre,Ivanov:2022qqt,Charalambous_2021,Kobayashi:2025vgl,Barack:2023oqp,Ivanov:2024sds,Caron-Huot:2025tlq,Combaluzier--Szteinsznaider:2025eoc,Kosmopoulos:2025rfj,Parra-Martinez:2025bcu} (but not in higher dimenions \cite{Kol:2011vg,Hui:2020xxx,Akhtar:2025nmt}).

A powerful such framework modeling compact objects with the point-particle approximation is the worldline quantum field theory (WQFT) \cite{Mogull:2020sak,Jakobsen:2021zvh,Jakobsen:2022psy,Haddad:2024ebn}.
Dimensional analysis indicates that the first spinless non-minimal operators which may deform the point-particle WQFT enter at $\cO(G^{4})$ \cite{Haddad:2024ebn}, translating to an $\cO(G^{6})$ contribution to two-body scattering.
Accurate modeling past the current 5PM state-of-the-art thus requires that these new parameters be included in the action and their specific black hole values identified.
This can be done uniquely only through a matching calculation to general relativity (GR).
The amplitude for a wave scattering off of a black hole (referred to in the literature as the ``Compton" or ``Raman" amplitude, though both names have connotations not applicable to our results) is an appropriate, gauge-invariant conduit for this matching, with the GR results deriving from black-hole perturbation theory (BHPT)~\cite{Mano:1996mf,aminov2020blackholequasinormalmodes,Bautista:2023sdf}.

Though single-body WQFT quantities have been considered relatively little compared to their two-body counterparts \cite{Ben-Shahar:2023djm,Haddad:2025cmw,He:2025how}, recent work has applied this framework to derive the amplitude mentioned above to NLO, but stopped short of comparing to BHPT \cite{Bjerrum-Bohr:2025bqg}.
In this paper, we push these results to NNLO ($\cO(G^{3})$),\footnote{Our NNLO $N$-matrix element is consistent with the result announced by Julio Parra-Martinez at the \emph{QCD meets Gravity 2024} conference~\cite{julioqcdmeetsgravity} and shared with us in private correspondence with Zihan Zhou.
 } and demonstrate the equivalence between the amplitude as computed in WQFT and BHPT.
While this precision is not yet sensitive to tidal effects, our main goal here is to develop the WQFT computation and BHPT matching, paving the way for higher-precision studies and Wilson-coefficient matching in the future.

In particular, we will advocate for working with the exponential representation of the $S$ matrix, $\hat{S}=e^{i\hat{N}}$.
In addition to better infrared and collinear behavior than the $T$-matrix elements usually defining amplitudes, we will show that the matrix element of $\hat{N}$ can be related directly to the scattering phase shift that one computes in BHPT.
This meshes nicely with the recent identification of the classical limit of this operator with the generator of classical scattering \cite{Gonzo:2024zxo,Kim:2024svw,Kim:2025olv,Kim:2025gis,Kim:2025hpn,Haddad:2025cmw,Alessio:2025flu}.
Nevertheless, the $T$-matrix element up to NNLO for this process is an intermediate step in our calculation of the $N$-matrix element.
Outside of the present context, the $T$-matrix element we produce can still be useful for the computation of other binary-black hole observables, such as the scattering waveform at higher loop orders \cite{Brandhuber:2023hhy,Brunello:2025eso,Brunello:2024ibk,Bini:2024rsy,Elkhidir:2023dco,Georgoudis:2023lgf,Georgoudis:2023ozp,Georgoudis:2024pdz,Georgoudis:2023eke}.
As such, we report it as well in the supplementary material.

Novel techniques for the calculation of black hole Love numbers have recently emerged \cite{Correia:2024jgr,Caron-Huot:2025tlq,Combaluzier--Szteinsznaider:2025eoc,Kosmopoulos:2025rfj}.
Despite the efficiency and innovation of these formalisms, one must be aware that superficially similar operators may need different coefficient values in different formalisms to reproduce the same observables: take, for example, $C_{E^{2}S^{4}}=0$ in ref.~\cite{Haddad:2024ebn} compared to $c_{E^{2}S^{4}}=-1$ in ref.~\cite{Ben-Shahar:2023djm}.\footnote{Another example is an open problem in the matching of the scattering self-force computation of ref.~\cite{Bini:2024icd} to the on-shell amplitudes results of ref.~\cite{Barack:2023oqp}, which produced
a non-vanishing static tidal contribution to the scattering angle of two compact objects with a scalar source at 4PM order.  
}
Additionally, the bypassing of the calculation of an amplitude makes translating results between different formalisms unwieldy.
For these reasons, matching the amplitude (rather than an effective wave equation) as derived from WQFT is a valuable endeavor.
Looking ahead, as spin effects are readily incorporated into the WQFT framework~\cite{Haddad:2024ebn}, this program may eventually enable the determination of the gravitational ``Compton'' amplitude for Kerr black holes beyond the spin-hexadecapole order, where spin-contact operators are known to mix with tidal deformations~\cite{Bautista:2024agp}; see also refs.~\cite{Dolan:2008kf,Bautista:2021wfy,Aoude:2022trd,Bern:2022kto,Bautista:2022wjf,Bautista:2023sdf,Akpinar:2025byi,Ben-Shahar:2025tiz,Kim:2025xka}.

In the remainder of the paper, we review the computation of the amplitude for the scattering of a gravitational wave off of a black hole in both WQFT and BHPT and present the strategy used to match the two approaches (\Cref{sec:ScatteringTheory}).
Then, we present our computation of the $N$-matrix element, from integrand generation and integration (\Cref{sec:Methods}) to the final result for the WQFT amplitude and comparison to BHPT (\Cref{sec:Results}).
Our conclusions are presented in \Cref{sec:conclusions}.
We provide the covariant $T$-matrix elements, the $N$-matrix elements in the frame of the BH, and the master integrals that we have used, all up to two-loop order, in the supplementary material.

In \Cref{app:npointfunctions}, we review the formalities of classical $n$-point functions in WQFT.
\Cref{app:Integrals} collects the results for the Feynman integrals involved up to NNLO, and \Cref{app:SWSH} summarizes pertinent properties of spin-weighted spherical harmonics.

\section{Wave scattering in WQFT and BHPT}\label{sec:ScatteringTheory}

In this section we review the computation of the scattering of a gravitational wave off of a black hole both in WQFT and in BHPT.
We also lay out our matching strategy, demonstrating that the exponential representation of the $S$-matrix produces amplitudes which map directly onto the BHPT phase shift.

\begin{figure}
  \includegraphics[width=.6\textwidth]{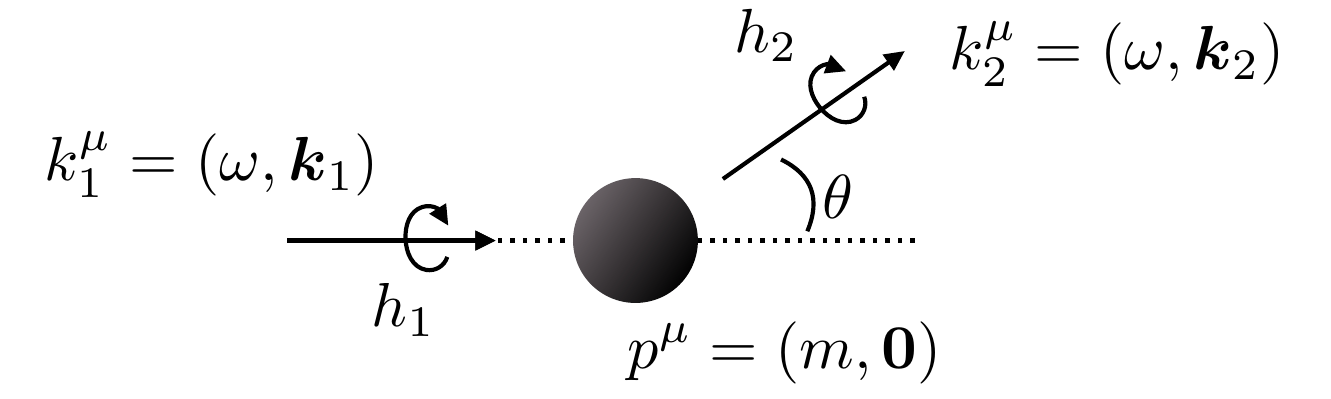}
  \centering
  \caption{
    Schematic of our kinematic setup.
    An idealized plane wave of definite momentum $k_1^\mu$ and helicity $h_1$ scatters off a black hole with momentum $p^\mu$.
    The scattered wave comes out at an angle $\theta$ with momentum $k_2^\mu$ and helicity $h_2$.
  }
  \label{fig:kinematics}
\end{figure}

Our kinematic setup is sketched in \cref{fig:kinematics}.
Traditionally, the  BHPT analysis for such a scattering process is done in the rest frame of the BH. 
In this frame, an idealized (i.e. spatially non-localized) gravitational plane wave with frequency $\omega$ impinges on a (spinless) black hole with mass $m$; it is customary to orient the $z$-axis along the direction of propagation of this incoming wave.
The helicity of the incoming wave may generally be reversed in the scattering process, in contrast to the scattering of a scalar or vector wave, and is due to graviton self-interactions \cite{Dolan:2007ut}.
After the scattering, an outgoing wave emerges with a generic orientation parametrized by the scattering angle $\theta$ and azimuthal angle $\phi$.
Due to the symmetry of the setup, the value of $\phi$ has no special significance, and indeed only enters amplitudes through an overall phase.

Parametrizing the black hole four-velocity as $v^\mu=p^\mu/m=(1,\boldsymbol{0})$, and the incoming and outgoing plane wave momenta and polarizations as
\begin{equation}\label{eq:Kinematics}
    \begin{aligned}
        k_{1}^{\mu}
        =
        \omega(1,0,0,1)
        ,&
        \quad k_{2}^{\mu}
        =
        \omega(1,s_\theta\,c_\phi,s_\theta\,s_\phi,c_\theta), 
        \\
        \epsilon_{1}^{\sigma_{1},\mu}
        =
        \frac{1}{\sqrt{2}}(0,1,i\sigma_{1},0)
        ,&
        \quad 
        \epsilon_{2}^{\sigma_{2},\mu}
        =
        \frac{1}{\sqrt{2}}\left(
          0,c_\theta\,c_\phi-i\sigma_{2}s_\phi,c_\theta\,s_\phi+i\sigma_{2}c_\phi,-s_\theta
          \right),
    \end{aligned}
\end{equation}
will align our WQFT calculations with available GR results.
Here, we use the shorthands $s_{y}\equiv\sin(y)$ and $c_{y}\equiv\cos(y)$, and $\sigma_i$ are the signs of the graviton helicities $h_i=2\sigma_i$.
In the helicity basis, the graviton polarization tensors are then $\eps^{\sigma_i,\mu\nu}_i=\eps^{\sigma_i,\mu}_i\eps^{\sigma_i,\nu}_i$.
The amplitudes presented later involve the complex conjugate of $\epsilon^{\sigma_{2},\mu\nu}_{2}$, because graviton 2 is outgoing.
The above gauge choice for $\epsilon_i^{\sigma_i,\mn}$ is commonly referred to as transverse traceless (TT) gauge and has the properties
\begin{align}\label{eq:TTGauge}
  \epsilon_i^{\sigma_i,\mn} v_\nu =0
  \ ,
  \quad
  \epsilon_i^{\sigma_i,\mn} \eta_\mn=0
  \ .
\end{align}
As a side note, the (circular) helicity basis chosen here is related to the linear plus-cross basis through $\epsilon_i^{\sigma_i,\mn}=\epsilon_i^{+,\mn}+\sigma_i i \epsilon_i^{\times,\mn}$.

In the following, it will often be useful to express the scattering angle $\theta$ in terms of the variable
\begin{align}\label{eq:x}
  x\equiv\sin(\theta/2),
\end{align}
which rationalizes expressions in the amplitude and integrals.
Moreover, the scattering will only depend on the relative helicities through the variable
\begin{align}\label{eq:HelicityProd}
  \sigma\equiv\sigma_1 \sigma_2 = \frac{h_1 h_2}{|h_1||h_2|}
  \ .
\end{align}
The values $\sigma=1$ or $\sigma=-1$ imply that helicity is preserved or reversed respectively.

This scattering process involves two scales: the frequency $\omega$ of the incoming (and outgoing) wave and the mass $m$ of the BH.
In this paper we consider the low-frequency regime of the scattering, in which the dimensionless expansion parameter
\begin{align}\label{eq:EpsPert}
  \epsPM\equiv 2G m \omega\,,
\end{align}
comparing the Schwarzschild radius of the black hole to the frequency of the incoming wave, is small.
Instead of the small frequency, one may interpret Newton's constant as a formal expansion parameter.
In this sense, our expansion is post-Minkowskian.
This identification persists for binary-black hole scattering, where the frequency of the exchanged waves is small (or, equivalently, their wavelength is of the order of a large impact parameter $b$, such that the expansion parameter becomes $Gm/b$).

\subsection{Scattering in WQFT}
WQFT employs an effective description of the black hole as a point-like particle~\cite{Goldberger:2004jt}, with its (spinless) degrees of freedom captured by a worldline field $x^\mu(\tau)$.
With the gravitational field and its dynamics described by the Einstein-Hilbert term, the total action reads
\begin{equation}\label{eq:WQFTAction}
    \begin{aligned}
        &\hspace{4.4cm}S
        =
        S_{\rm wl}+S_{\rm EH}+S_{\rm gf}
        \ , \\
        &S_{\rm wl}
        =
        -m\int{\rm d}\tau\,\tfrac{1}{2}
        g_{\mu\nu}(x)\,
        \dot{x}^{\mu}(\tau)\dot{x}^{\nu}(\tau)
        \ ,
        \quad
        S_{\rm EH}
        =
        -\frac1{16\pi G}
        \int{\rm d}^{D}x\,\sqrt{-g}R
        \ ,
    \end{aligned}
\end{equation}
where the worldline action is in ``Polyakov'' form~\cite{DESER1976369,Brink:1976sc,Mogull:2020sak} and we have added a de-Donder type gauge fixing term $S_{\rm gf}$.
Gravitational dynamics are kept in arbitrary dimensions,
\begin{align}
  D=4-2\epsDR
  \ ,
\end{align}
in order to dimensionally regularize integrals.

Perturbation theory is established by expanding all fields around their incoming ``background'' behavior:
\begin{align}\label{eq:backgroundExpansion}
  g_\mn(x)
  = 
  \eta_\mn
  +
  h^{(0)}_\mn(x)
  +
  \kappa h_\mn(x)
  \ ,
  \qquad
  x^\mu(\tau)
  =
  \tau v^\mu
  +
  z^\mu(\tau)
  \ ,
\end{align}
where $\kappa=\sqrt{32\pi G}$.
We have put the black hole at the spatial origin, ignoring a possible impact parameter $b^\mu$ in its background parametrization.
The background value of the gravitational field $h^{(0)}_\mn(x)$ describes the incoming wave, and must obey the equations of motion of linearized gravity (here in de-Donder type gauge):
\begin{align}
   \pat^\alpha \pat_\alpha h_\mn^{(0)}(x)=0
   \ , \qquad 
   \pat^\mu \mathcal{P}_\mn^\ab h_\ab^{(0)}(x)=0
   \ ,\qquad
   \mathcal{P}_\mn^\ab
   =
   \eta^{(\alpha}_\mu\eta^{\beta)}_\nu
   -\tfrac12 \eta^\ab \eta_\mn
   \ .
\end{align}
Namely, in the general case, $h^{(0)}_\mn(x)$ is a superposition of plane gravitational waves.
In particular, a single gravitational plane wave with definite helicity $h_0=2\sigma_0$ and momentum $k_0$ has the form
\begin{align}
  h^{(0)}_\mn(x;k_0)
  =e^{-ik_0\cdot x}\epsilon^{\sigma_0}_{\mn}(k_0)
  =\int \frac{{\rm d}^{D}k}{(2\pi)^{D}} e^{-ik\cdot x}
  \dd^D(k-k_0) \epsilon^{\sigma_0}_\mn(k_0)
  \ .
\end{align}
Here, we introduced our conventions for transforming to momentum space.
Factors of $2\pi$ are absorbed into $\dd(x)\equiv2\pi\,\delta(x)$.

The WQFT framework solves for asymptotic observables in terms of the background variables through a Feynman diagrammatic expansion, in which causality is imposed via retarded propagators, formally following from Schwinger-Keldysh in-in theory~\cite{Jakobsen:2022psy,Kalin:2022hph}.
Explicitly, we work in momentum space and represent the two dynamical fields, the graviton $h_\mn(k)$ and the deflection $z^\mu(\omega)$, by wavy and solid lines respectively:
\begin{align}
  \begin{tikzpicture}[baseline=-1mm]
    \gravitonLine{(0,0)}{(1.2,0)}{.6}
    \node[below] at (.6,-.1) {$k$} ;
    \node[left] at (0,0) {$\mu\nu$} ;
    \node[right] at (1.2,0) {$\alpha\beta$} ;
  \end{tikzpicture}
  =
  i\frac{\mathcal{P}^{-1}_{\mn\ab}}{(k^0+i0^+)^2-\mathbf{k}^2}
  \ ,
  \quad
  \begin{tikzpicture}[baseline=-1mm]
    \zLine{(0,0)}{(1.2,0)}
    \node[below] at (.6,-.1) {$\omega$} ;
    \node[left] at (0,0) {$\mu$} ;
    \node[right] at (1.2,0) {$\alpha$} ;
  \end{tikzpicture}
  =
  -i\frac{\eta^{\mu\alpha}}{m(\omega+i0^+)^2}
  \ .
\end{align}
The incoming wave gives rise to a one-point vertex rule for the graviton which we depict as\footnote{Note that $h^{(0)}_\ab(k)$ must be kept slightly off-shell for this Feynman vertex to be non-zero. One may take the on-shell limit after contracting with a propagator.
}
\begin{equation}\label{eq:incomingWave}
    \begin{tikzpicture}[baseline=-1mm]
    \draw [photonTest] (0,0) -- (1.2,0) ;
    \node[below] at (.7,-.1) {$k$} ;
    \node[right] at (1.2,0) {$\mu\nu$} ;

    \draw [fill=white,draw=white] (0,0) circle (0.3cm);
    \begin{scope}[rotate=45]
      \clip (0,0) circle (0.3cm);
      \foreach \y in {-0.6,-0.5,...,0.6} {
        \draw[
          decorate,
          decoration={snake, amplitude=1pt, segment length=6pt}
        ] (-1,\y) -- (1,\y);
      }
    \end{scope}

  \end{tikzpicture}
  =\,
  -i k^2 \mathcal{P}^{\mn\ab} h^{(0)}_\ab(k)
  \,\underset{\text{plane wave}}{=}\,
  -i k^2 \mathcal{P}^{\mn\ab}
  \dd^D(k-k_0)\epsilon^{\sigma_0}_\ab(k)
  \ .
\end{equation}
This rule cannot simply be derived by plugging the background expansion eq.~\eqref{eq:backgroundExpansion} into the Einstein-Hilbert action.
Instead, it is best understood in connection with the graviton propagator.
Together they invert the kinetic term of the graviton: the vertex rule in eq.~\eqref{eq:incomingWave} imposes the boundary conditions in the infinite past (an incoming wave) and the retarded propagator ensures that no other waves are present initially.
Properly speaking, then, this vertex rule applies to the total flat-space perturbation $h_\mn^{(0)}(x)+\kappa h_\mn(x)$, though the difference will not matter in this work.

Interactions between the graviton and the black hole are well studied, and are described by worldline interactions~\cite{Mogull:2020sak,Jakobsen:2023oow,Haddad:2025cmw}.
In principle, an arbitrary number of worldline fluctuations $z^{\mu}$ can interact with one graviton, although in this work we only need the first two of this tower of vertices:
\begin{equation}\label{eq:FeynmanRulesWL}
\begin{aligned}
  \begin{tikzpicture}[baseline=15pt]
    \draw [photonTest] (0,0) -- (0,1.2) ;
    \node[right] at (.1,.7) {$k$} ;
    \node[above] at (0,1.1) {$\mu\nu$} ;
    \draw [dotted, thick] (-.5,0) -- (.5,0) ;
    \draw [fill] (0,0) circle (.08);
  \end{tikzpicture}
  &=
  -i\frac{m\kappa}{2}
   \dd(k\cdot v) v^\mu v^\nu 
  \ , \\
  \begin{tikzpicture}[baseline=15pt]
    \draw [photonTest] (0,0) -- (0,1.2) ;
    \node[right] at (.1,.7) {$k$} ;
    \node[above] at (0,1.1) {$\mu\nu$} ;
    \draw [dotted, thick] (.5,0) -- (0,0) ;
    \draw [wl] (-.7,0) -- (0,0) ;
    \node[below] at (-.4,-0.1) {$\omega$} ;
    \node[left] at (-.7,0) {$\sigma$} ;
    \draw [fill] (0,0) circle (.08);
  \end{tikzpicture}
  &=
  \frac{m\kappa}{2}
   \dd(k\cdot v+\omega)
  (
    v^\mu v^\nu k_\sigma
    +2\omega v^{(\mu}\eta^{\nu)}_{\sigma}
  )
  \ .
\end{aligned}
\end{equation}
Here, all momenta are outgoing.
Finally, gravitational self-interactions are described by $n$-graviton vertices well-known from the Einstein-Hilbert action.

The outgoing gravitational waveform describes the metric perturbation emerging from a gravitational interaction.
In WQFT, it is computed as the on-shell graviton one-point function $\langle h_\mn(k)\rangle$.
More precisely, the expansion of the position-space waveform at spatial infinity reads~\cite{Jakobsen:2021smu,Jakobsen:2021lvp,Cristofoli:2021vyo}
\begin{align}\label{eq:waveform}
  \epsilon^{*\mn} 
  h_\mn(x)
  \underset{|\vct{x}|\to\infty}{\to}
  -\frac{1}{4\pi |\vct{x}|}
  \int \frac{{\rm d}\omega}{2\pi} 
  e^{-i\omega (t-|\vct{x}|)}
  k^2 
  \epsilon^{*\mn}(k) 
  \big\langle
  h_\mn(k)
  \big\rangle
  \Big|_{k=\omega(1,\hat{\vct{x}})}.
\end{align}
We contract $h_\mn$ with the complex conjugate $\epsilon^{*\mn}$ because the field is outgoing.
To say more, we must specify the process that sources the outgoing wave.

In this work we focus on scattering effects that are linear in the incoming wave (i.e. ``linear'' black hole perturbation theory).
For example, the leading-order 1PM (i.e. $\cO(\epsPM)$) diagrams read\footnote{We note that the computation of these diagrams was carried out explicitly without spin in ref.~\cite{Bjerrum-Bohr:2025bqg} and to quadratic order in spins as a check for ref.~\cite{Saketh:2022wap}.}
\begin{align}\label{eq:TreeLevel}
  \big\langle h_\mn(k)\big\rangle\Big|_{\epsPM^1}
  \ \ 
  =\ \ 
  \begin{tikzpicture}[baseline=15pt]
    \gravitonLine{(0,0)}{(0,1.2)}{.6}
    \gravitonLine{(-1,1.2)}{(0,1.2)}{.7}
    \gravitonLine{(0,1.2)}{(1,1.2)}{.6}
    \node[below] at (.7,1.2) {$k$} ;
    \node[right] at (1,1.2) {$\mu\nu$} ;
    \draw [dotted, thick] (-.5,0) -- (.5,0) ;
    \snakeblob{(-1,1.2)}{45}
    \draw [fill] (0,0) circle (.08);
    \draw [fill] (0,1.2) circle (.08);
  \end{tikzpicture}
  \ \ +\ \ 
  \begin{tikzpicture}[baseline=15pt]
    \gravitonLine{(-1,1.2)}{(0,0)}{.7}
    \gravitonLine{(1,0)}{(2,1.2)}{.6}
    \draw [dotted, thick] (-.5,0) -- (0,0) ;
    \draw [dotted, thick] (1,0) -- (1.5,0) ;
    \node[below] at (1.9,1.) {$k$} ;
    \node[right] at (1.65,1.4) {$\mu\nu$} ;
    \draw [zParticle] (0,0) -- (1,0) ;
    \snakeblob{(-1,1.2)}{45}
    \draw [fill] (0,0) circle (.08);
    \draw [fill] (1,0) circle (.08);
  \end{tikzpicture}.
\end{align}
More generally, this process may conveniently be related to a graviton two-point function,
\begin{align}\label{eq:LinBHPTWF}
  -ik^2 
  \epsilon^{*\mn}(k) 
  \big\langle h_\mn(k)\big\rangle
  \ \ =\ \ 
  \begin{tikzpicture}[baseline=4mm]
    \draw [dotted, thick] (-1.5,0) -- (1.5,0) ;
    \draw [photonTest] (0,0) -- (-1.2,1.2) ;
    \draw [example1={.5}]  (-1.2,1.2) -- (0,0) ;
    \draw [photonTest] (0,0) -- (1.2,1.2) ;
    \draw [example1={.7}] (0,0) -- (1.2,1.2) ;
    \node[below] at (1.2,1.) {$k$} ;
    \node[right] at (0.95,1.4) {$\epsilon^\mn(k)$} ;
    \draw [fill=black!60, thick] (0,0.1) circle (0.4cm);
    \snakeblob{(-1.2,1.2)}{45}
  \end{tikzpicture}
  \ ,
\end{align}
where the grey blob (borrowing notation from ref.~\cite{Bjerrum-Bohr:2025bqg}) represents the connected graviton two-point function $\langle h_\mn(k) h_\ab(k_0)\rangle$. 
This in turn is amputated by the on-shell polarization and momentum factor explicit on the left-hand side of this equation and the background vertex in \cref{eq:incomingWave}.
In other words, \cref{eq:LinBHPTWF} represents the transfer matrix of the scattering process connecting the incoming wave with the outgoing one,
\begin{align}
  -ik^{2}\epsilon^{*\mn}(k)\langle h_{\mn}(k)\rangle=\langle h(k)|i\hat{T}|h^{(0)}(k_0)\rangle.
\end{align}
Specializing to idealized plane waves, this reduces to the $T$-matrix element familiar from QFT (though below we will only consider classically relevant diagrams; more in \cref{sec:Methods}):
\begin{equation}\label{eq:DefTMatrixElement}
  \langle k_2,h_2|i \hat T| k_1,h_1\rangle
  \,\,=\,\, 
  \begin{tikzpicture}[baseline=4mm]
    \draw [dotted, thick] (-1.5,0) -- (1.5,0) ;
    \draw [photonTest] (0,0) -- (-1.2,1.2) ;
    \draw [example1={.5}]  (-1.2,1.2) -- (0,0) ;
    \draw [photonTest] (0,0) -- (1.2,1.2) ;
    \draw [example1={.7}] (0,0) -- (1.2,1.2) ;
    \node[right] at (0.75,1.45) {$k_2,h_2$} ;
    \node[left] at (-0.75,1.45) {$k_1,h_1$} ;
    \draw [fill=black!60, thick] (0,0.1) circle (0.4cm);
  \end{tikzpicture}
  \,\,=\,\, 
  \dd(k_2\cdot v-k_1\cdot v)
  \,i
  \cM_{\sigma}(\theta,\phi).
\end{equation}
Here we have explicitly chosen a basis of definite helicities $h_i$ and incoming and outgoing momenta $k_1^\mu$ and $k_2^\mu$ respectively, as introduced in \cref{eq:Kinematics} and depicted in \cref{fig:kinematics}.
The final equality of \cref{eq:DefTMatrixElement} defines the amplitude $\cM_\sigma(\theta,\phi)$ characterizing the scattering, and its arguments refer to the explicit kinematics of \cref{eq:Kinematics}.
This amplitude is what is commonly referred to as the gravitational ``Compton" amplitude.
We refrain from using this terminology here as it traditionally carries quantum mechanical connotations.

Combining \cref{eq:waveform,eq:DefTMatrixElement} (and thus still assuming an incoming idealized plane wave), the outgoing spherical wave takes the form
\begin{align}\label{eq:WaveformAmplitude}
  \epsilon_2^{*\sigma_{2},\mn} h_\mn(x)
  \underset{|\vct{x}|\to\infty}{\to}
  \frac{1}{4\pi|\vct{x}|}
  e^{-i(t-|\vct{x}|)\omega}
  \cM_{\sigma}(\theta,\phi)
  \ ,
\end{align}
Here, $\vct{k}_2$ is parallel to $\vct{x}$ according to eq.~\eqref{eq:waveform}.
In other words, the angles $\theta$ and $\phi$ entering the scattering amplitude are the spherical coordinates of $\vct{x}$.

Taking stock of our discussion so far, we have understood the outgoing gravitational waveform in terms of a scattering amplitude.
While this analysis reflects the theoretical and computational underpinnings of our approach, the scattering amplitude itself, as defined through \cref{eq:DefTMatrixElement}, does not optimally encode the relevant classical information.
One indication of this is its poor infrared behavior, due to the long-range nature of the gravitational interaction.
Rather than the waveform, BHPT analyses package the scattering information in the infrared finite phase shift, reviewed in the next section.
It is therefore in our interest to seek the most direct path from the scattering amplitude so far considered to the BHPT phase shift.

The transfer matrix is canonically related to the scattering (or $S$) matrix through $\hat{S}=1+i\hat{T}$.
On the other hand, the $S$ matrix may be made to resemble a phase shift by recasting it in an exponential representation \cite{Damgaard:2021ipf,Brandhuber:2025igz},
\begin{align}\label{eq:SMatrixExp}
    \hat{S}=e^{i\hat{N}}.
\end{align}
In recent works, the classical limit of the operator $\hat{N}$ -- also referred to as the \emph{Magnus operator}
or \emph{Magnusian} \cite{Magnus1954,BLANES2009151,Ebrahimi_Fard_2025} -- has been identified with the generator of classical observables \cite{Kim:2024svw,Kim:2025gis,Kim:2025olv,Kim:2025hpn} (see also \cite{Kosower:2018adc,Parra-Martinez:2020dzs,Bern:2021dqo,Kol:2021jjc,Alessio:2025flu,Haddad:2025cmw}).
In a similar vein, beyond the superficial resemblance, we will see shortly that matrix elements of $\hat{N}$ are indeed closely related to the classical BHPT phase shift.
In particular, $N$-matrix elements are infrared finite and even exponentiate in the appropriate representation for the process under consideration.\footnote{$N$-matrix elements are also free of classically-singular contributions in the binary scattering setting \cite{Damgaard:2021ipf,Haddad:2025cmw}.}
Unitarity of the $S$-matrix implies that $N$-matrix elements are real (up to little group rotations).

Absent a direct mechanism for computing the $N$-matrix element, we will pass through the $T$ matrix as an intermediate quantity.
Defining $\hat{X}_{fi}\equiv\langle f|\hat{X}|i\rangle$, equating both expressions for the $S$ matrix leads to relations between the matrix elements of $\hat{N}$ and $\hat{T}$: expanding $\langle f|\log\hat{S}|i\rangle$ to NNLO in a perturbative coupling gives
\begin{subequations}\label{eq:NFromT}
\begin{align}
    i\hat{N}_{fi}^{(1)}&=i\hat{T}_{fi}^{(1)},\label{eq:TtoN1} \\
    i\hat{N}_{fi}^{(2)}&=i\hat{T}_{fi}^{(2)}-\frac{1}{2}\left(i\hat{T}^{(1)}\cdot i\hat{T}^{(1)}\right)_{fi},\label{eq:TtoN2} \\
    i\hat{N}_{fi}^{(3)}&=i\hat{T}_{fi}^{(3)}-\frac{1}{2}\left(i\hat{T}^{(2)}\cdot i\hat{T}^{(1)}+i\hat{T}^{(1)}\cdot i\hat{T}^{(2)}\right)_{fi}+\frac{1}{3}\left(i\hat{T}^{(1)}\cdot i\hat{T}^{(1)}\cdot i\hat{T}^{(1)}\right)_{fi},\label{eq:TtoN3}
\end{align}
\end{subequations}
Practically, the products of $T$ matrices are inner products on the Hilbert space, involving the insertion of a complete set of states.
Though this is a quantum mechanical concept, restricting to the classical context of this work is straightforward, entailing only single-graviton states:
\begin{align}
  \hat X \cdot \hat Y
  &=
  \sum_{h=\pm}\int\frac{{\rm d}^{D}\ell}{(2\pi)^{D}}\dd(\ell^{2})
  \hat X|\ell,h\rangle
  \langle\ell,h|
  \hat Y
  \ .
\end{align}
In a quantum mechanical setting, one would have to consider further contributions from multi-particle states.

\Cref{eq:NFromT} thus allows us to obtain $N$-matrix elements from $T$-matrix elements.
After reviewing the calculation of wave scattering in GR, we will see that this also grants us immediate access to the BHPT phase shift.

\subsection{Scattering in BHPT}

On the GR side, the amplitude for a wave scattering off of a Schwarzschild black hole is extracted from the    
$r\to\infty$ form of the solution to the homogenous spin-weight $2$ Regge-Wheeler \cite{PhysRev.108.1063} and Zerilli \cite{PhysRevLett.24.737} equations, respectively controlling the axial and polar modes of the gravitational perturbation.

In modern languge, the curvature perturbation can be packaged in the radiation scalar of spin-weight $s=-2$, namely  $r^{4}{}_{-2}\psi_4(r,\theta,\phi,t)$, which satisfies a second order partial differential equation known as the Prince-Bardeen-Press-Teukolsky equation \cite{PhysRevD.5.2419,Bardeen1973,Teukolsky1973}. 
The spherical symmetry of the problem allows for the projection of the curvature perturbation onto a spherical basis,\footnote{Since this work is restricted to spinless black holes, the kinematic conventions adopted for the incoming wave in \cref{eq:Kinematics} fix the azimuthal mode number to $m = 2$. This should not be confused with the black hole mass, which is also denoted by $m$ throughout this work.
 }
\begin{align}\label{eq:CurvaturePertSeparated}
  r^{4}{}_{-2}\psi_4(r,\theta,\phi,t) \sim \sum_{\ell} e^{-i\omega t}{}_{-2}Y_{\ell 2}(\theta,\phi) {}_{-2}\psi_{\ell 2}(r),
\end{align}
which separates the PDE into an angular component, solved by spin-weighted spherical harmonics ${}_{-2}Y_{\ell m}(\theta,\phi)$ \cite{BHPToolkit}, and a radial component, solved by ${}_{-2}\psi_{\ell 2}(r)$.

The desired long-distance behavior is ultimately controlled by the radial ODE, which is an example of a confluent Heun differential equation \cite{Ronveaux1995}.
In particular, the infrared-finite partial-wave scattering matrix ${}_{-2}S^P_{\ell 2}$ can be read off from the asymptotic form of the scalar solution ${}_{-2}\psi_{\ell 2}(r)$,
\begin{equation}\label{eq:CurvaturePertAsymptotic}
  {}_{-2}\psi_{\ell 2}(r)\overset{r^{\star}\rightarrow\infty}{\longrightarrow}\frac{e^{-i \omega r^\star}}{\omega r}+ {}_{-2}S^P_{\ell 2} \frac{e^{i \omega r^\star}}{(\omega r)^5}\,,
\end{equation}
where $P=\pm1$ is a parity label and $r^\star = r+ 2 Gm \log(\frac{r}{2G m}-1)$ is the tortoise coordinate.
It is also customary to decompose the  scattering matrix  into a real elastic phase shift ${}_{-2}^{\vphantom{P}}\delta_{\ell 2}^{P}$ and a real absorption factor ${}_{-2}\eta_{\ell 2}$ as
\begin{equation}
  {}_{-2}S^P_{\ell 2} =  {}_{-2}\eta_{\ell2} e^{2 i {}_{-2}^{\vphantom{P}}\delta_{\ell 2}^{P}}.
\end{equation} 
Subtracting the ``no interaction'' scenario, the helicity-preserving and helicity-reversing scattering amplitudes are respectively
given by the gravitational analog of the Rayleigh-Faxen-Holtsmark formula as~\cite{Dolan:2007ut}
\begin{subequations}\label{eq:AmplitudeBHPT}
\begin{align}
  f(\theta,\phi) &=
  \frac{\pi}{\omega}\sum_{\ell=2}^\infty \sqrt{\frac{2\ell+1}{\pi}}{}_{-2}Y_{\ell 2}(\theta,\phi) \sum_{P=\pm1} \left({}_{-2}\eta_{\ell2}e^{2i{}_{-2}\delta_{\ell 2}^{P}}-1\right)\,,\label{eq:ThpBHPT}\\
  g(\theta,\phi) &=
  \frac{\pi}{\omega}\sum_{\ell=2}^\infty \sqrt{\frac{2\ell+1}{\pi}}{}_{-2}Y_{\ell 2}(\pi-\theta,\phi) \sum_{P=\pm1} (-1)^\ell P\left({}_{-2}\eta_{\ell2}e^{2i{}^{\vphantom{P}}_{-2\vphantom{\ell}}\delta_{\ell 2}^{P}}-1\right)\label{eq:ThrBHPT}\,.
\end{align}
\end{subequations} 
Both the elastic phase shift and absorption factor admit a perturbative expansion in $\epsPM$.
The latter obeys ${}_{-2}\eta_{\ell2}=1+\cO(\epsPM^{5})$, so will not be relevant at the $\cO(\epsPM^{3})$ precision we compute in this paper.
The amplitudes in \cref{eq:AmplitudeBHPT} are computable using either the MST method \cite{Mano:1996mf}, or Nekrasov-Shatashvili functions \cite{Bautista:2023sdf}.

\subsection{Matching strategy}\label{sec:Matching}

To connect the discussions of the preceding two sections, one must translate the waveform in \cref{eq:WaveformAmplitude} to (the outgoing part of) the curvature scalar in \cref{eq:CurvaturePertAsymptotic}.
A relevant discussion on this procedure can be found in ref.~\cite{Bautista:2021wfy}.
We will not engage further with the details here, but rather only state the relation between the WQFT and BHPT scattering amplitudes:
\begin{align}\label{eq:RelateAmplitudes}
  \cM_{1}(\theta,\phi)\overset{!}{=}2\pi f(\theta,\phi)\,,\quad \cM_{-1}(\theta,\phi)\overset{!}{=}2\pi g(\theta,\phi)\,.
\end{align}
Instead of matching at the level of the amplitudes, we will do so directly at the level of the phase shift.

Specifically, we will prove that, up to the precision considered in this paper,
\begin{subequations}\label{eq:deltalWQFT}
  \begin{align}
    \frac{1}{4\pi}\frac{\omega}{\sqrt{\pi(2\ell+1)}}\int{\rm d}\Omega_{2}\, {}_2 Y_{\ell, -2}(\theta,\phi) \bar{N}^{(n)}_{2,2}(\omega;\theta,\phi) = & \sum_{P=\pm1} {}_{-2}\delta_{\ell 2}^{P,(n)}\,,\label{eq:deltalWQFThp}\\ 
    \frac{1}{4\pi}\frac{\omega}{\sqrt{\pi(2\ell+1)}}\int{\rm d}\Omega_{2}\, {}_2 Y_{\ell,- 2}(\pi-\theta,\phi) \bar{N}^{(n)}_{2,-2}(\omega;\theta,\phi) = & \sum_{P=\pm1} P(-1)^\ell {}_{-2}\delta_{\ell 2}^{P,(n)}\label{eq:deltalWQFThr}\,,
  \end{align}
\end{subequations}
where ${}_{-2}\delta_{\ell2}^{P,(n)}$ is the $n$\textsuperscript{th} term in the low-frequency expansion of the BHPT scattering  phase shift,
\begin{align}
  \langle k_{2},h_{2}|\hat{N}|k_{1},h_{1}\rangle\big|_{\epsPM^{n}}=\dd(v\cdot k_{1}-v\cdot k_{2})N^{(n)}_{h_{1},h_{2}}(\omega;\theta,\phi),
\end{align}
and $\bar{N}_{h_{1},h_{2}}^{(n)}$ is a regularized amplitude to be defined below.
In order to prove this we will show that, in a certain sense, the matrix element of $\hat{N}$ can be exponentiated.

To begin, we will consider the matrix element with the most general graviton momenta, $q_{i}^{\mu}$ for $i=1,2$, but still in the rest frame of the worldline, $v^{\mu}=(1,\boldsymbol{0})$:
\begin{align}
  \langle q_{2},h_{2}|\hat{N}|q_{1},h_{1}\rangle=\dd(v\cdot q_{1}-v\cdot q_{2})N_{h_{1},h_{2}}(\omega;\alpha,\beta,\gamma).
\end{align}
The energy of the incoming wave $\omega=v\cdot q_{1}$ is conserved by the Dirac delta function on the right-hand side.
Each graviton four-momentum is parametrized by two angles as
\begin{align}
  q_{i}^{\mu}&=\omega(1,\sin\theta_{i}\,\cos\phi_{i},\sin\theta_{i}\,\cos\phi_{i},\cos\theta_{i}).
\end{align}
While one may write the matrix element $N_{h_{1},h_{2}}$ in terms of these angles directly, it suits our purposes to instead study this element as a function of the Euler angles
\begin{subequations}\label{eq:EulerAngles}
  \begin{align}
    \cot\alpha&=\cos\theta_{2}\cot(\phi_{2}-\phi_{1})-\cot\theta_{1}\sin\theta_{2}\csc(\phi_{2}-\phi_{1})\,,\label{eq:Eulera} \\
    \cos\beta&=\cos\theta_{1}\cos\theta_{2}+\sin\theta_{1}\sin\theta_{2}\cos(\phi_{2}-\phi_{1})\,,\label{eq:Eulerb} \\
    \cot\gamma&=\cos\theta_{1}\cot(\phi_{2}-\phi_{1})-\cot\theta_{2}\sin\theta_{1}\csc(\phi_{2}-\phi_{1})\,.\label{eq:Eulerg}
  \end{align}
\end{subequations}
Considering the little group weight of the gravitational matrix element and the possible scalar products which may appear in the rest frame, its most general form at any loop order in terms of Euler angles is
\begin{align}\label{eq:AmplitudeEuler}
  N_{h_{1},h_{2}}(\omega;\alpha,\beta,\gamma)=e^{i(h_{1}\gamma+h_{2}\alpha)}\frac{1}{\omega}\sum_{n=0}^{2}\frac{F_{n}(\cos\beta)}{(\cos\beta-\sigma)^{n}}\,.
\end{align}
All physical poles of the amplitude in $\beta$ are poles of the form factors $F_{n}$, while those apparent here originate from the Lorentz products $\epsilon^{\sigma_{1},\mu}(q_{1})\left[\epsilon^{\sigma_{2}}_{\mu}(q_{2})\right]^{*}$ and are therefore spurious.
The significance of this decomposition is that all helicity dependence has been made explicit in terms of $h_{i}$ and $\sigma$ (defined in \cref{eq:HelicityProd}), and the $F_{n}$ specifically are helicity-independent.
More details about the Euler angles and this decomposition can be found in \Cref{app:EulerAngles}.
Assuming that the $F_{n}$ are analytic in their argument, we may split them into even and odd components
\begin{align}\label{eq:AmplitudeAnalytic}
  F_{n}(z)=F_{n}^{+}(z)+F_{n}^{-}(z),
\end{align}
where $F_{n}^{\pm}(-z)=\pm F_{n}^{\pm}(z)$.
This assumption of analyticity holds at least up to two-loop order, and we will discuss below when we expect it to fail.

Now, to make contact with the phase shift from BHPT, we decompose the matrix element onto a basis of spin-weighted spherical harmonics,
\begin{align}\label{eq:AmplitudeDecomposition}
  N_{h_{1},h_{2}}(\omega;\alpha,\beta,\gamma)=\frac{1}{\omega}\sum_{\ell=2}^{\infty}e^{ih_{2}\alpha}\,{}_{-h_{2}}Y_{\ell h_{1}}(\beta,\gamma)(-1)^{\ell(1-\sigma)/2}N_{\ell;h_{1},h_{2}},
\end{align}
where the dimensionless modes $N_{\ell;h_{1},h_{2}}$ can be extracted by using the orthonormality properties of the spin-weighted spherical harmonics, \cref{eq:SWSHOrtho}.
Doing so, exploiting the general decomposition \eqref{eq:AmplitudeEuler}, the analyticity assumption \eqref{eq:AmplitudeAnalytic}, and properties of the harmonics in \Cref{app:SWSH}, we can show that
\begin{align}
  N_{\ell;h_{1},h_{2}}&=2\pi\,(N_{\ell;+}+\sigma N_{\ell;-}),
\end{align}
where
\begin{align}
  N_{\ell;\pm}&\equiv e^{i h_1 \gamma}\int_{-1}^{1}{\rm d}z\,{}_{h_{1}}Y_{\ell(-h_{1})}(\arccos z,\gamma)\left[F_{0}^{\pm}(z)-\frac{F_{1}^{\mp}(z)}{1-z}+\frac{F_{2}^{\pm}(z)}{(1-z)^{2}}\right],
\end{align}
for $z=\cos\beta$.

Strictly speaking, \cref{eq:AmplitudeDecomposition} is only valid if the amplitude is integrable over the 2-sphere.
This is the case for all but the leading-order helicity-preserving $N$-matrix element.\footnote{The $T$-matrix elements are not integrable on the sphere at any order in perturbation theory.
This is because of the exponentiation of infrared singularities, which carry the tree-level pole to all orders.}
A rigorous approach therefore necessitates projecting on $(D-2)$-dimensional basis functions, as was done in refs.~\cite{Caron-Huot:2022jli,Ivanov:2024sds}, however we show now that this individual case can be adequately accommodated without introducing a higher-dimensional formalism.

At leading order, the graviton propagator produces a quadratic pole in the forward limit --  $\theta\rightarrow0$ in \cref{eq:Kinematics} -- of the helicity-preserving amplitude.\footnote{This divergence has an analog on the on the BHPT side, which we discuss further in \Cref{sec:Results}.}
Consequently, the inversion
\begin{align}
  N_{\ell;h,h}^{(1)}=\omega\,e^{-ih\alpha}\int{\rm d}\Omega\,{}_{-h}\bar{Y}_{\ell,h}(\beta,\gamma)N_{h,h}^{(1)}(\omega;\alpha,\beta,\gamma)
\end{align}
does not exist.
We can assign a value to this integral by promoting the measure to $2-2\varepsilon$ dimensions:
\begin{equation}\label{eq:TreeLevelModes}
  \begin{aligned}
    N_{\ell;h,h}^{\varepsilon,(1)}&\equiv \omega\,e^{-ih\alpha}\left(\frac{\bar{\mu}}{\omega}\right)^{2\varepsilon}\int{\rm d}\Omega^{(2-2\varepsilon)}\,{}_{-h}\bar{Y}_{\ell,h}(\beta,\gamma)N_{h,h}^{(1)}(\omega;\alpha,\beta,\gamma) \\
    &=\omega\,e^{-ih\alpha}\left(\frac{\bar{\mu}}{\omega}\right)^{2\varepsilon}\int{\rm d}\Omega\,\frac{\left(\pi\sin\beta\right)^{-2\varepsilon}}{\Gamma(1-\varepsilon)}{}_{-h}\bar{Y}_{\ell,h}(\beta,\gamma)N_{h,h}^{(1)}(\omega;\alpha,\beta,\gamma),
  \end{aligned}
\end{equation}
with some infrared scale $\bar{\mu}$.
The regularized mode\footnote{An inherently $D$-dimensional approach would have the regularization of this mode built in, but the divergence would still arise.
This is because its origin is the improper handling of long-range forces in the $S$-matrix itself. See refs.~\cite{Hannesdottir:2019opa,Lippstreu:2025jit,DeAngelis:2025vlf} for recent explorations.}
\begin{align}\label{eq:TreeLevelReg}
  \frac{\bar{N}_{\ell;h,h}^{(1)}}{\sqrt{\pi(2\ell+1)}}
  =\lim_{\varepsilon\rightarrow0}\left[\frac{N_{\ell;h,h}^{\varepsilon,(1)}}{\sqrt{\pi(2\ell+1)}}+4\pi\frac{\epsPM}{\varepsilon}\right]
\end{align}
is finite and turns out to reproduce the leading-order phase shift from BHPT.
As such, the regularized projection
\begin{align}\label{eq:AmplitudeDecompositionReg}
  \bar{N}_{h_{1},h_{2}}(\omega;\alpha,\beta,\gamma)=\frac{1}{\omega}\sum_{\ell=2}^{\infty}e^{ih_{2}\alpha}\,{}_{-h_{2}}Y_{\ell h_{1}}(\beta,\gamma)(-1)^{\ell(1-\sigma)/2}\bar{N}_{\ell;h_{1},h_{2}},
\end{align}
is the appropriate object to match.
The definite-parity modes $N_{\ell;\pm}$ are regularized according to
\begin{align}
  \frac{\bar{N}_{\ell;\pm}}{\sqrt{\pi(2\ell+1)}}
  =\lim_{\varepsilon\rightarrow0}\left[\frac{N^{\varepsilon}_{\ell;\pm}}{\sqrt{\pi(2\ell+1)}}+2\pi\frac{\epsPM}{\varepsilon}\right].
\end{align}
Again, only the $\cO(\epsPM)$ modes are modified.

The direct matrix element $\langle k_{2},h_{2}|\hat{N}|k_{1},h_{1}\rangle$ (i.e. in the frame \eqref{eq:Kinematics}) is the particular configuration of \cref{eq:AmplitudeDecomposition} where $(\theta_{1},\phi_{1})=(0,0)$ and $(\theta_{2},\phi_{2})=(\theta,\phi)$, giving $\alpha=0$, $\beta=\theta$, $\gamma=\phi$.
We can now make concrete the exponentiation of the amplitude to which we have alluded above: it is the modes $\bar{N}_{\ell;\pm}$ for the external states $\langle k_{2},h_{2}|$ and $|k_{1},h_{1}\rangle$ which exponentiate in the element of the $S$-matrix.
Let us be schematic about the rest of the argument, as the specifics are not particularly enlightening.

Additionally to the direct matrix element, exponentiation requires us to understand matrix elements of multiple insertions of $\hat{N}$.
The simplest such case is
\begin{align}
  \langle k_{2},h_{2}|\hat{N}\cdot\hat{N}|k_{1},h_{1}\rangle=\sum_{h=\pm2}\int\frac{{\rm d}^{D}q}{(2\pi)^{D}}\,\dd(q^{2})\langle k_{2},h_{2}|\hat{N}|q,h\rangle\langle q,h|\hat{N}|k_{1},h_{1}\rangle.
\end{align}
As the matrix elements contain energy-conserving delta functions, the energy integral is trivial.
Subsequently, the delta function setting $q^{\mu}$ on shell enables integrating out its radial component.
We are left with an integral over the solid angle of $\boldsymbol{q}$,
\begin{align}\label{eq:DoubleMEAngularIntegral}
  \langle k_{2},h_{2}|\hat{N}\cdot\hat{N}|k_{1},h_{1}\rangle\sim\sum_{h}\int{\rm d}\Omega^{(D-2)}_{\boldsymbol{q}}\,N_{h,h_{2}}(\omega;\alpha_{k_{2}|q},\beta_{k_{2}|q},\gamma_{k_{2}|q})N_{h_{1},h}(\omega;0,\theta_{q},\phi_{q}),
\end{align}
where the subscript $k_{2}|q$ indicates an Euler angle composed of the four angles $(\theta,\phi)$ and $(\theta_{q},\phi_{q})$ parametrizing $\boldsymbol{k}_{2}$ and $\boldsymbol{q}$.
Decomposing the matrix elements according to \cref{eq:AmplitudeDecompositionReg} allows us to set $D=4$, and isolates the angular dependence in spin-weighted spherical harmonics.
At this point, the so-called addition theorem, \cref{eq:SWSHAdd}, enables us to perform the final integrations.
Once the dust has settled, this matrix element has a mode decomposition of the form
\begin{align}
  \langle k_{2},h_{2}|\hat{N}\cdot\hat{N}|k_{1},h_{1}\rangle\big|_{\rm analytic}^{\rm reg.}\sim\sum_{\ell}\frac{{}_{-h_{2}}Y_{\ell h_{1}}(\theta,\phi)}{\omega\sqrt{2\ell+1}}(-1)^{\ell(1-\sigma)/2}\left(\bar{N}_{l;+}^{2}+\sigma\bar{N}_{l,-}^{2}\right),
\end{align}
where we remind the reader that we consider the regulated, analytic part of the matrix element.
The crucial difference to \cref{eq:AmplitudeDecomposition} is that the modes $\bar{N}_{\ell;\pm}$ here appear squared.

One can repeat the calculation for more insertions of $\hat{N}$, finding that the pattern continues.
Summing up all insertions, then, gives
\begin{subequations}\label{eq:AmplitudeExponentiated}
  \begin{align}
    &\langle k_{2},2|e^{i\hat{N}}-1|k_{1},2\rangle\big|_{\rm analytic}^{\rm reg.} \\
    &=2\pi\dd(v\cdot k_{1}-v\cdot k_{2})\frac{\pi}{\omega}\sum_{\ell=2}^{\infty}\,{}_{-2}Y_{\ell 2}(\theta,\phi)\sqrt{\frac{2\ell+1}{\pi}}\sum_{P=\pm1}\left[\exp\left(\frac{i\bar{N}_{\ell;P}}{\sqrt{\pi(2\ell+1)}}\right)-1\right],\nn \\
    &\langle k_{2},-2|e^{i\hat{N}}-1|k_{1},2\rangle\big|_{\rm analytic}^{\rm reg.} \\
    &=2\pi\dd(v\cdot k_{1}-v\cdot k_{2})\frac{\pi}{\omega}\sum_{\ell=2}^{\infty}\,{}_{2}Y_{\ell 2}(\pi-\theta,\phi)\sqrt{\frac{2\ell+1}{\pi}}\sum_{P=\pm1}P\left[\exp\left(\frac{i\bar{N}_{\ell;P}}{\sqrt{\pi(2\ell+1)}}\right)-1\right].\nn
  \end{align}
\end{subequations}
Note that the operator whose expectation value is being taken is $i\hat{T}$, so that the left-hand sides of these equations are simply $\dd(v\cdot k_{1}-v\cdot k_{2})\cM_{\sigma}(\theta,\phi)$.
Comparing the modes of \cref{eq:AmplitudeBHPT,eq:AmplitudeExponentiated} via \cref{eq:RelateAmplitudes}, it is now immediate to show
\begin{subequations}
  \begin{align}
    \frac{1}{4\pi}\frac{\bar{N}_{\ell;2,2}^{(n)}}{\sqrt{\pi(2\ell+1)}}&=\sum_{P=\pm}{}_{-2}\delta_{\ell 2}^{P,(n)}, \\
    \frac{1}{4\pi}\frac{\bar{N}_{\ell;2,-2}^{(n)}}{\sqrt{\pi(2\ell+1)}}&=\sum_{P=\pm}P(-1)^{\ell}\,{}_{-2}\delta_{\ell 2}^{P,(n)}.
  \end{align}
\end{subequations}
Plugging in the modes as projections of the amplitude, one finally arrives at \cref{eq:deltalWQFT}.

To close this section, let us briefly note that the exponential structure of the BHPT amplitude \eqref{eq:AmplitudeBHPT} is spoiled by the absorption factor ${}_{-2}\eta_{\ell 2}=1+\cO(\epsPM^{5})$.
Absorptive effects can be incorporated into the WQFT computation following refs.~\cite{Caron-Huot:2025tlq,Ivanov:2024sds,Kosmopoulos:2025rfj}, which should introduce non-analyticities in $\cos\beta$ to the $N$-matrix element.

\section{Diagrammatics and integration}\label{sec:Methods}

\begin{figure}[t]
  \centering
  \subcaptionbox{}{
    \begin{tikzpicture}
      \coordinate (currentLocation) at (0,0) ;
      \drawLtoLshort
      \drawLzero
      \drawLtoLshort
      \drawLzero
      \drawLtoLshort
    \end{tikzpicture}
  }\hfill
  \subcaptionbox{}{
    \begin{tikzpicture}
      \coordinate (currentLocation) at (0,0) ;
      \drawLtoLshort
      \drawLoneA
      \drawLtoLshort
    \end{tikzpicture}
  }\hfill
  \subcaptionbox{}{
    \begin{tikzpicture}
      \coordinate (currentLocation) at (0,0) ;
      \drawLtoLshort
      \drawLoneB
      \drawLtoLshort
    \end{tikzpicture}
  }\hfill
  \subcaptionbox{}{
    \begin{tikzpicture}
      \coordinate (currentLocation) at (0,0) ;
      \drawLtoLshort
      \drawLzero
      \drawLtoMshort
      \drawWL
      \drawMtoLshort
    \end{tikzpicture}
  }\hfill
  \subcaptionbox{}{
    \begin{tikzpicture}
      \coordinate (currentLocation) at (0,0) ;
      \drawLtoMshort
      \drawWL
      \drawMtoLshort
      \drawLzero
      \drawLtoLshort
    \end{tikzpicture}
  }\hfill
  \subcaptionbox{}{
    \begin{tikzpicture}
      \coordinate (currentLocation) at (0,0) ;
      \drawLtoMshort
      \drawWL
      \drawM
      \drawWL
      \drawMtoLshort
    \end{tikzpicture}
  }
  \caption{
    The six one-loop diagrams.
    Using retarded propagators, all causality flows from left to right.
    Only active (red) gravitons can go on-shell and the $i0^+$ prescription matters only for those.
    Graphs (4)-(6) vanish in the TT gauge implied by \cref{eq:Kinematics}.
    The six graphs have the symmetry factors (1, $1/2$, $1/2$, 1, 1, 1).
  }
  \label{fig:oneLoop}
\end{figure}
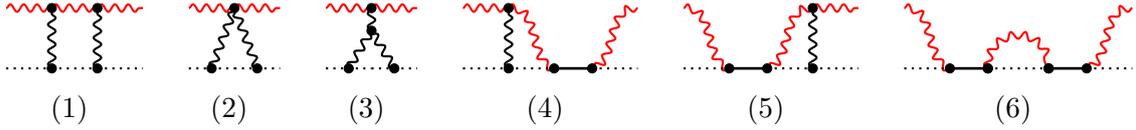
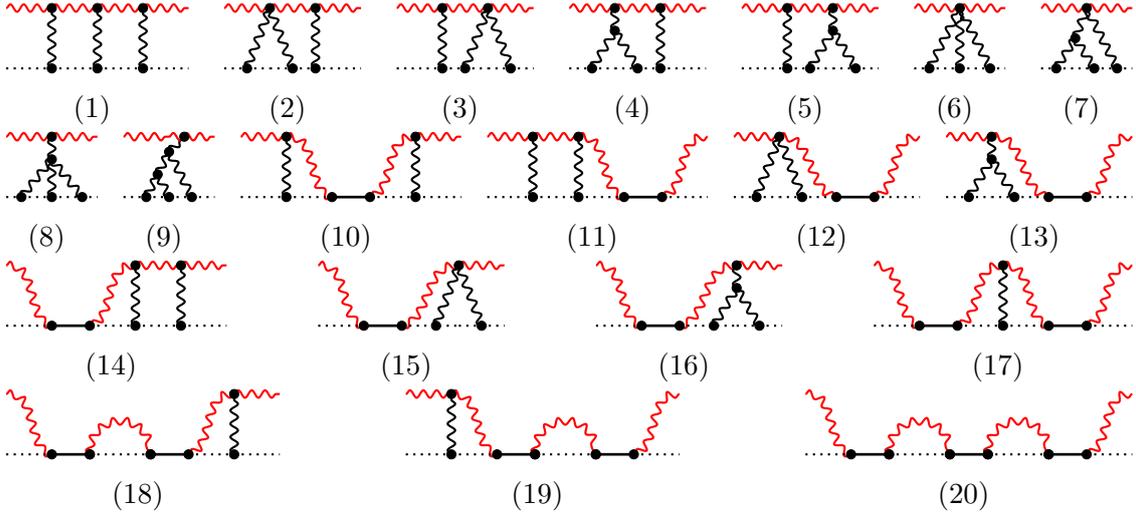
\begin{figure}[t]
  \centering
  \subcaptionbox{}{
    \begin{tikzpicture}
      \coordinate (currentLocation) at (0,0) ;
      \drawLtoLshort
      \drawLzero
      \drawLtoLshort
      \drawLzero
      \drawLtoLshort
      \drawLzero
      \drawLtoLshort
    \end{tikzpicture}
  }\hfill
  \subcaptionbox{}{
    \begin{tikzpicture}
      \coordinate (currentLocation) at (0,0) ;
      \drawLtoLshort
      \drawLoneA
      \drawLtoLshort
      \drawLzero
      \drawLtoLshort
    \end{tikzpicture}
  }\hfill
  \subcaptionbox{}{
    \begin{tikzpicture}
      \coordinate (currentLocation) at (0,0) ;
      \drawLtoLshort
      \drawLzero
      \drawLtoLshort
      \drawLoneA
      \drawLtoLshort
    \end{tikzpicture}
  }\hfill
  \subcaptionbox{}{
    \begin{tikzpicture}
      \coordinate (currentLocation) at (0,0) ;
      \drawLtoLshort
      \drawLoneB
      \drawLtoLshort
      \drawLzero
      \drawLtoLshort
    \end{tikzpicture}
  }\hfill
  \subcaptionbox{}{
    \begin{tikzpicture}
      \coordinate (currentLocation) at (0,0) ;
      \drawLtoLshort
      \drawLzero
      \drawLtoLshort
      \drawLoneB
      \drawLtoLshort
    \end{tikzpicture}
  }\hfill
  \subcaptionbox{}{
    \begin{tikzpicture}
      \coordinate (currentLocation) at (0,0) ;
      \drawLtoLshort
      \drawLtwoA
      \drawLtoLshort
    \end{tikzpicture}
  }\hfill
  \subcaptionbox{}{
    \begin{tikzpicture}
      \coordinate (currentLocation) at (0,0) ;
      \drawLtoLshort
      \drawLtwoB
      \drawLtoLshort
    \end{tikzpicture}
  }\hfill
  \subcaptionbox{}{
    \begin{tikzpicture}
      \coordinate (currentLocation) at (0,0) ;
      \drawLtoLshort
      \drawLtwoC
      \drawLtoLshort
    \end{tikzpicture}
  }\hfill
  \subcaptionbox{}{
    \begin{tikzpicture}
      \coordinate (currentLocation) at (0,0) ;
      \drawLtoLshort
      \drawLtwoD
      \drawLtoLshort
    \end{tikzpicture}
  }\hfill
    \subcaptionbox{}{
  \begin{tikzpicture}
    \coordinate (currentLocation) at (0,0) ;
    \drawLtoLshort
    \drawLzero
    \drawLtoMshort
    \drawWL
    \drawMtoLshort
    \drawLzero
    \drawLtoLshort
  \end{tikzpicture}
}\hfill
  \subcaptionbox{}{
    \begin{tikzpicture}
      \coordinate (currentLocation) at (0,0) ;
      \drawLtoLshort
      \drawLzero
      \drawLtoLshort
      \drawLzero
      \drawLtoMshort
      \drawWL
      \drawMtoLshort
    \end{tikzpicture}
  }\hfill
  \subcaptionbox{}{
    \begin{tikzpicture}
      \coordinate (currentLocation) at (0,0) ;
      \drawLtoLshort
      \drawLoneA
      \drawLtoMmedium
      \drawWL
      \drawMtoLshort
    \end{tikzpicture}
  }\hfill
  \subcaptionbox{}{
    \begin{tikzpicture}
      \coordinate (currentLocation) at (0,0) ;
      \drawLtoLshort
      \drawLoneB
      \drawLtoMmedium
      \drawWL
      \drawMtoLshort
    \end{tikzpicture}
  }\hfill
  \subcaptionbox{}{
  \begin{tikzpicture}
    \coordinate (currentLocation) at (0,0) ;
    \drawLtoMshort
    \drawWL
    \drawMtoLshort
    \drawLzero
    \drawLtoLshort
    \drawLzero
    \drawLtoLshort
  \end{tikzpicture}
}\hfill
  \subcaptionbox{}{
  \begin{tikzpicture}
    \coordinate (currentLocation) at (0,0) ;
    \drawLtoMshort
    \drawWL
    \drawMtoLmedium
    \drawLoneA
    \drawLtoLshort
  \end{tikzpicture}
}\hfill
  \subcaptionbox{}{
  \begin{tikzpicture}
    \coordinate (currentLocation) at (0,0) ;
    \drawLtoMshort
    \drawWL
    \drawMtoLmedium
    \drawLoneB
    \drawLtoLshort
  \end{tikzpicture}
}\hfill
  \subcaptionbox{}{
  \begin{tikzpicture}
    \coordinate (currentLocation) at (0,0) ;
    \drawLtoMshort
    \drawWL
    \drawMtoLshort
    \drawLzero
    \drawLtoMshort
    \drawWL
    \drawMtoLshort
  \end{tikzpicture}
}\hfill
  \subcaptionbox{}{
  \begin{tikzpicture}
    \coordinate (currentLocation) at (0,0) ;
    \drawLtoMshort
    \drawWL
    \drawM
    \drawWL
    \drawMtoLshort
    \drawLzero
    \drawLtoLshort
  \end{tikzpicture}
}\hfill
  \subcaptionbox{}{
  \begin{tikzpicture}
    \coordinate (currentLocation) at (0,0) ;
    \drawLtoLshort
    \drawLzero
    \drawLtoMshort
    \drawWL
    \drawM
    \drawWL
    \drawMtoLshort
  \end{tikzpicture}
}\hfill
  \subcaptionbox{}{
  \begin{tikzpicture}
    \coordinate (currentLocation) at (0,0) ;
    \drawLtoMshort
    \drawWL
    \drawM
    \drawWL
    \drawM
    \drawWL
    \drawMtoLshort
  \end{tikzpicture}
}
  \caption{
    The 20 two-loop diagrams.
    Using retarded propagators, all causality flows from left to right.
    Only active (red) gravitons can go on-shell and the $i0^+$ prescription matters only for those.
    In TT gauge, only graphs (1)-(10) are non-zero.
    Graphs (2, 3, 4, 5, 7, 9, 12, 13, 15, 16) have symmetry factor 1/2, graphs (6, 8) have symmetry factor $1/6$ and all other graphs have symmetry factor 1.
  }
  \label{fig:twoLoop}
\end{figure}

We present here some of the details of our WQFT calculation. 
We focus in particular on the generation of the relevant diagrams, as well as providing an instructive overview of loop-integration techniques.

\subsection{Integrand generation}\label{subsec:integrand}

All diagrams contributing to the graviton two-point function $\langle h_\mn(x_1) h_\ab(x_2)\rangle$ may be derived recursively through a Berends-Giele-like relation, which itself emerges from the classical terms in the ``quantum action'' of WQFT.
We include this analysis in \Cref{app:npointfunctions}, and the relevant relation is \cref{eq:EOMForTwoPoints}.

Essentially, this equation manifests the fact that the graviton two-point function is a (curved space) Green's function of a certain differential operator.
Up to ``recoil terms'' \cite{Cheung:2023lnj}, this operator is the so-called Einstein operator (see, e.g., ref.~\cite{Pound:2021qin}) of the black hole metric, which has the diagrammatic representation
\begin{equation}
    \begin{tikzpicture}[baseline=-6mm,xscale=-1]

    \draw [dotted, thick] (-1., -0.7) -- (1, -0.7);

    \draw [photonTest] (0,-0.4) -- (1.,-0.4);
    \draw [example1={.65}] (1.6,-0.4) -- (0,-0.4) ;
    \draw [photonTest] (0,-0.4) -- (-1,-0.4);
    \draw [example1={.65}] (0,-0.4) -- (-1,-0.4);
    \draw [fill=white, thick] (0,-0.6) circle (0.3cm);
    \node at (0,-0.6) {\Large$\times$};
  \end{tikzpicture}
  \,\,
  =
  \,\,
  \sum_{n=1}^\infty\ \frac1{n!}\,\,
  \begin{tikzpicture}[baseline=0mm,xscale=-1]
    \draw [dotted, thick] (-1.5, -0.4) -- (1.5, -0.4);

    \draw [photonTest] (0,0.6) -- (-1,0.6);
    \draw [example1={.55}] (0,0.6) -- (-1,0.6);
    \draw [photonTest] (0,0.6) -- (1,0.6);
    \draw [example1={.55}] (1,0.6) -- (0,0.6) ;
    \draw [fill] (0,0.6) circle (.08);
    \draw [photonTest] (0,0.6) -- (-0.6,-0.4);
    \draw [fill=black!60, thick] (-0.6,-0.4) circle (0.25cm);
    \draw [photonTest] (0,0.6) -- (0.6,-0.4);
    \draw [fill=black!60, thick] (0.6,-0.4) circle (0.25cm);
    \node at (-.04,-0.2) {$\cdots$};
    \node at (0.03,-.9) {$\underbrace{\hphantom{hej hej hej}}_n$};
  \end{tikzpicture}
  \ .
\end{equation}
Concretely, the grey blob with a single outgoing graviton is the non-perturbative graviton one-point function in the absence of incoming waves, which is precisely the background Schwarzschild metric.
Building on this, a rearrangement of \cref{eq:EOMForTwoPoints} has the diagrammatic representation
\begin{equation}\label{eq:BerendsGiele}
  \begin{tikzpicture}[baseline=1mm,xscale=-1]
    \draw [dotted, thick] (-1,0) -- (0,0);
    \draw [photonTest] (0,0) -- (-1,.8);
    \draw [example1={.65}] (0,0) -- (-1,.8);
    \draw [photonTest] (1,.8) -- (0,0);
    \draw [example1={.5}] (1,.8) -- (0,0);
    \draw [dotted, thick] (-1,0) -- (1,0);
    \drawCompton{(0,0)}
  \end{tikzpicture}
  \ \ =\ \ 
  \begin{tikzpicture}[baseline=1mm,xscale=-1]
    \draw [dotted, thick] (0,0) -- (1.2,0);
    \draw [photonTest] (1.2,.3) -- (0,.3);
    \draw [example1={.6}] (1.2,.3) -- (0,.3);
  \end{tikzpicture}
  \ \ 
  +\ \ 
  \begin{tikzpicture}[baseline=1mm,xscale=-1]

    \draw [dotted, thick] (-.6,0) -- (1.2,0);
    \draw [photonTest] (0.4,0) -- (-.6,.8);
    \draw [example1={.55}] (0.4,0) -- (-.6,.8);

    \draw [zParticle] (1.2,0) -- (0.4,0);
    \draw [fill] (0.4,0) circle (.08);
    \draw [fill] (1.2,0) circle (.08);
    \draw [photonTest] (1.2,0) to[out=80, in=100] (2.4,0);
    \draw [example1={.65}] (2.4,0) to[out=100, in=80] (1.2,0) ;
    \draw [photonTest] (2.4,0) -- (3.4,.8);
    \draw [example1={.5}] (3.4,.8) -- (2.4,0);
    \draw [dotted, thick] (1.2,0) -- (3.4,0);
    \drawCompton{(2.4,0.1)}
  \end{tikzpicture}
  \ \ +\ \ 
  \begin{tikzpicture}[baseline=-6mm,xscale=-1]

    \draw [dotted, thick] (-.6, -0.7) -- (2.6, -0.7);

    \draw [photonTest] (2.6,0.2) -- (1.6,-0.6);
    \draw [example1={.5}] (2.6,0.2) -- (1.6,-0.6);
    \draw [photonTest] (0.4,-0.4) -- (1.8,-0.4);
    \draw [example1={.6}] (1.8,-0.4) -- (0.4,-0.4) ;
    \draw [photonTest] (0.4,-0.6) -- (-.6,0.2);
    \draw [example1={.6}] (0.4,-0.6) -- (-.6,0.2);
    \drawCompton{(1.6,-.6)}
    \draw [fill=white, thick] (0.4,-0.6) circle (0.3cm);
    \node at (0.4,-0.6) {\Large$\times$};
  \end{tikzpicture}
  \ \ .
\end{equation}
In this equation, one must lift the external legs of the graviton two point function (grey blob) from on-shell plane waves to generic off-shell momenta.

One may solve the Berends-Giele relation~\eqref{eq:BerendsGiele} recursively for the graviton two-point function.
For example, its (off-shell) $\epsPM^0$ behavior is the flat space propagator which, when inserted on the right-hand-side of \cref{eq:BerendsGiele} gives rise to the tree-level $\epsPM^1$ terms of \cref{eq:TreeLevel}.
The recursive relation is reminiscent of a Schwinger-Dyson self-energy relation.
As such, the graviton two-point function admits a formal resummation in terms of a geometric series:
\begin{equation}
  \begin{tikzpicture}[baseline=1mm,xscale=-1]
    \draw [dotted, thick] (-1,0) -- (1,0);
    \draw [photonTest] (0,0) -- (-1,.8);
    \draw [example1={.65}] (0,0) -- (-1,.8);
    \draw [photonTest] (1,.8) -- (0,0);
    \draw [example1={.5}] (1,.8) -- (0,0);
    \drawCompton{(0,0)}
  \end{tikzpicture}
  \ \ =\ \ 
  \frac{
    \begin{tikzpicture}[baseline=1mm,xscale=-1]
    \draw [dotted, thick] (0,0) -- (1.2,0);
    \draw [photonTest] (1.2,.3) -- (0,.3);
    \draw [example1={.6}] (1.2,.3) -- (0,.3);
  \end{tikzpicture}
  }{
  1\ \ -\ \ 
  \begin{tikzpicture}[baseline=2mm,xscale=-1]

    \draw [dotted, thick] (-.8,0) -- (1.0,0);
    \draw [photonTest] (0.2,0) -- (-.8,.8);
    \draw [example1={.55}] (0.2,0) -- (-.8,.8);

    \draw [zParticle] (1.,0) -- (0.2,0);
    \draw [fill] (0.2,0) circle (.08);
    \draw [fill] (1,0) circle (.08);
    \draw [photonTest] (1,0) -- (2.,.8);
    \draw [example1={.65}] (2.,.8) -- (1,0) ;
    \draw [dotted, thick] (1,0) -- (2,0);
  \end{tikzpicture}
  \ \ -\ \ 
  \begin{tikzpicture}[baseline=-4mm,xscale=-1]

    \draw [dotted, thick] (-1., -0.6) -- (1, -0.6);

    \draw [photonTest] (0,-0.6) -- (1.,0.2);
    \draw [example1={.6}] (1.,0.2) -- (0,-0.6) ;
    \draw [photonTest] (0,-0.6) -- (-1,0.2);
    \draw [example1={.6}] (0,-0.6) -- (-1,0.2);
    \draw [fill=white, thick] (0,-0.6) circle (0.3cm);
    \node at (0,-0.6) {\Large$\times$};
  \end{tikzpicture}
  }
  \ \ .
\end{equation}
This provides intuition to the diagrammatic structure of the graviton two-point function.
Finally, we note that one might also have used the Berends-Giele relation for the graviton one-point function (in the presence of an incoming wave) as presented in refs.~\cite{Jakobsen:2023ndj,Jakobsen:2023oow}.
This, however, would have additionally produced diagrams for non-linear wave scattering which are not relevant for this work.

Applying our diagram-generating formula~\eqref{eq:BerendsGiele}, we find two diagrams at tree level, which were given in \cref{eq:TreeLevel}.
Continuing to one and two loops we find a total of 6 and 20 diagrams respectively, drawn in \cref{fig:oneLoop,fig:twoLoop}.
In these diagrams, gravitons of different colors possess different causal properties.
Black gravitons are \textit{potential}: their energy components in the black hole frame $v^\mu=(1,\vct{0})$ vanish.
On the other hand, red gravitons are \textit{active}: their energy components are fixed to $\omega$.
Both graviton types have spatial components which scale as $\omega$, such that potential gravitons are space-like while active gravitons may go on shell.
This reflects our physical setup, in which an initially on-shell graviton travels through the potential of a BH.
We note that with the TT gauge choice~\eqref{eq:TTGauge} only half of the diagrams at one and two loops are non-zero.

Insert the Feynman rules~\eqref{eq:FeynmanRulesWL} into our diagrams completes the construction of our integrand, making integration the only remaining step before obtaining the amplitude.

\subsection{Integration}\label{subsec:integration}
From the little group scaling of the external gravitons, we deduce that the maximal tensor rank of integrals at any loop order is four.
Performing a Passarino--Veltman reduction \cite{Passarino:1978jh}, tensor integrals are reduced to scalar ones, all of which fit into the $L$-loop family
\begin{subequations}\label{eq:IntegralFamilies}
\begin{align}
  \int \d^{D} \ell_1 \ldots \d^{D} \ell_L \frac{\prod_{n=1}^{L}\hat{\delta}(v \cdot (\ell_n - k_1))}{\prod_{n=1}^{N} D_n^{\nu_n}},
\end{align}
where the $N=3L+{L\choose2}$ denominators are
\begin{align}
  D_{n}\in\{(\ell_i^0 + i0^+)^2 - \vct{\ell}_i^2,(\ell_i-k_1)^2,(\ell_i-k_2)^2,(\ell_i-\ell_j)^2\}
\end{align}
for $1\leq i\neq j\leq L$.
\end{subequations}
One notable feature here as opposed to integrals for binary scattering is that worldline propagators do not appear, because loops for this process with such propagators are classically subleading in WQFT.
Consequently, energy conservation localizes the energy of any deflection mode to that of the external gravitons.

In the rest frame of the worldline all delta functions can be resolved to impose $\ell_i^0-k_j^0=0$ on the loop momenta.
Doing so leaves us with purely spatial integrals of the form
\begin{align}
  \int \d^{D-1} \vct{\ell}_1 \ldots \d^{D-1} \vct{\ell}_L \frac{1}{\prod_{n=1}^{N} \vct{D}_n^{\nu_n}},
\end{align}
where the spatial propagators are
\begin{align}
  \vct{D}_n \in \{
  \vct{\ell}_i^2-(\omega+i0^+)^2,(\vct{\ell}_i-\vct{k}_1)^2,(\vct{\ell}_i-\vct{k}_2)^2,(\vct{\ell}_i-\vct{\ell}_j)^2\}.
\end{align}
Integration-by-parts (IBP) reduction to a set of master integrals is simpler to carry out in this frame, since otherwise one may have to seed integrals with modified powers of the delta functions.
At two-loop order the IBPs are not challenging, but may become so at the higher loop orders we intend to undertake.
We perform the IBP reduction using \texttt{Kira 3} \cite{Lange:2025fba}.

At one-loop order we find three master integrals given diagrammatically as
\begin{align}\label{eq:1LoopMasters}
  \vec{I}_{\text{1-loop}}=\{I_1,I_2,I_3\}=\bigg\{\ \ 
  \begin{tikzpicture}[baseline=7pt]
    \coordinate (cm3) at (-.5,.65) ;
    \coordinate (cm1) at (-.5,0) ;
    \coordinate (cm4) at (+1.3,.65) ;
    \coordinate (cm2) at (+1.3,0) ;
    \coordinate (c3) at (+0,.65) ;
    \coordinate (c1) at (+0,0) ;
    \coordinate (c4) at (+.8,.65) ;
    \coordinate (c2) at (+.8,0) ;
    \coordinate (c5) at (+.4,.65) ;
    \coordinate (c6) at (+.4,.5) ;
    \draw[photonTest,red] (cm3) -- (c1) ;
    \draw[photonTest,red] (c2) -- (cm4) ;
    \draw[photonTest,red] (c1) to[out=90,in=180] (c6) to[out=0,in=90] (c2) ;
    \draw[dottedLine] (cm1) -- (cm2) ;
    \vertexSmall{(c1)}
    \vertexSmall{(c2)}
  \end{tikzpicture}
  \ \ ,\ \ 
  \begin{tikzpicture}[baseline=7pt]
    \draw[photonTest,red] (cm3) -- (c5) ;
    \draw[photonTest,red] (c5) -- (cm4) ;
    \draw[photonTest] (c1) -- (c5) ;
    \draw[photonTest] (c2) -- (c5) ;
    \draw[dottedLine] (cm1) -- (cm2) ;
    \vertexSmall{(c1)}
    \vertexSmall{(c2)}
    \vertexSmall{(c5)}
  \end{tikzpicture}
  \ \ ,\ \ 
  \begin{tikzpicture}[baseline=7pt]
    \draw[photonTest,red] (cm3) -- (c3) ;
    \draw[photonTest,red] (c3) -- (c4) ;
    \draw[photonTest,red] (c4) -- (cm4) ;
    \draw[photonTest] (c1) -- (c3) ;
    \draw[photonTest] (c2) -- (c4) ;
    \draw[dottedLine] (cm1) -- (cm2) ;
    \vertexSmall{(c1)}
    \vertexSmall{(c2)}
    \vertexSmall{(c3)}
    \vertexSmall{(c4)}
  \end{tikzpicture}
  \ \ 
  \bigg\}
  \ .
\end{align}
The six master integrals at two-loop order are
\begin{align}\label{eq:2LoopMasters}
    &\vec{I}_{\text{2-loop}}=\{I_1,I_2,I_3,I_4,I_5,I_6\}
    =\bigg\{\, 
    \begin{tikzpicture}[baseline=7pt]
      \coordinate (cm3) at (-.45,.65) ;
      \coordinate (cm1) at (-.45,0) ;
      \coordinate (cm4) at (+2.05,.65) ;
      \coordinate (cm2) at (+2.05,0) ;
      \coordinate (c3) at (+0,.65) ;
      \coordinate (c1) at (+0,0) ;
      \coordinate (c4) at (+.8,.65) ;
      \coordinate (c2) at (+.8,0) ;
      \coordinate (c5) at (+.4,.65) ;
      \coordinate (c6) at (+.4,.5) ;
      \coordinate (c8) at (+1.6,.65) ;
      \coordinate (c7) at (+1.6,0) ;
      \coordinate (c9) at (+1.2,.65) ;
      \coordinate (c10) at (+1.2,.5) ;
      \draw[photonTest,red] (cm3) -- (c1) ;
      \draw[photonTest,red] (c7) -- (cm4) ;
      \draw[photonTest,red] (c1) to[out=90,in=180] (c6) to[out=0,in=90] (c2) ;
      \draw[photonTest,red] (c2) to[out=90,in=180] (c10) to[out=0,in=90] (c7) ;
      \draw[dottedLine] (cm1) -- (cm2) ;
      \vertexSmall{(c1)}
      \vertexSmall{(c2)}
      \vertexSmall{(c7)}
    \end{tikzpicture}
    \,,\,
    \begin{tikzpicture}[baseline=7pt]
      \draw[photonTest,red] (c3) -- (c4) ;
      \draw[photonTest,red] (c8) -- (c4) ;
      \draw[photonTest] (c1) -- (c4) ;
      \draw[photonTest] (c2) -- (c4) ;
      \draw[photonTest] (c7) -- (c4) ;
      \draw[dottedLine] (cm1) -- (cm2) ;
      \vertexSmall{(c1)}
      \vertexSmall{(c2)}
      \vertexSmall{(c7)}
      \vertexSmall{(c4)}
    \end{tikzpicture}
    \,,\,
    \\
    &\hspace{2cm}
    \begin{tikzpicture}[baseline=7pt]
      \draw[photonTest,red] (cm3) -- (c5) ;
      \draw[photonTest,red] (c5) to[out=0,in=130] (c7) ;
      \draw[photonTest,red] (c7) -- (cm4) ;
      \draw[photonTest] (c1) -- (c5) ;
      \draw[photonTest] (c2) -- (c5) ;
      \draw[dottedLine] (cm1) -- (cm2) ;
      \vertexSmall{(c1)}
      \vertexSmall{(c2)}
      \vertexSmall{(c7)}
      \vertexSmall{(c5)}
    \end{tikzpicture}
    \,,\,
    \begin{tikzpicture}[baseline=7pt]
      \draw[photonTest,red] (cm3) -- (c5) ;
      \draw[photonTest,red] (c5) -- (c8) ;
      \draw[photonTest,red] (c8) -- (cm4) ;
      \draw[photonTest] (c1) -- (c5) ;
      \draw[photonTest] (c2) -- (c5) ;
      \draw[photonTest] (c7) -- (c8) ;
      \draw[dottedLine] (cm1) -- (cm2) ;
      \vertexSmall{(c1)}
      \vertexSmall{(c2)}
      \vertexSmall{(c7)}
      \vertexSmall{(c5)}
      \vertexSmall{(c8)}
    \end{tikzpicture}
    \,,\,
    \mathcal{N}\times
    \begin{tikzpicture}[baseline=7pt]
      \draw[photonTest,red] (cm3) -- (c3) ;
      \draw[photonTest,red] (c3) -- (c4) ;
      \draw[photonTest,red] (c4) -- (c8) ;
      \draw[photonTest,red] (c8) -- (cm4) ;
      \draw[photonTest] (c1) -- (c3) ;
      \draw[photonTest] (c2) -- (c4) ;
      \draw[photonTest] (c7) -- (c8) ;
      \draw[dottedLine] (cm1) -- (cm2) ;
      \vertexSmall{(c1)}
      \vertexSmall{(c2)}
      \vertexSmall{(c7)}
      \vertexSmall{(c3)}
      \vertexSmall{(c4)}
      \vertexSmall{(c8)}
    \end{tikzpicture}
    \,,\,
    \begin{tikzpicture}[baseline=7pt]
      \draw[photonTest,red] (cm3) -- (c3) ;
      \draw[photonTest,red] (c3) -- (c4) ;
      \draw[photonTest,red] (c4) -- (c8) ;
      \draw[photonTest,red] (c8) -- (cm4) ;
      \draw[photonTest] (c1) -- (c3) ;
      \draw[photonTest] (c2) -- (c4) ;
      \draw[photonTest] (c7) -- (c8) ;
      \draw[dottedLine] (cm1) -- (cm2) ;
      \vertexSmall{(c1)}
      \vertexSmall{(c2)}
      \vertexSmall{(c7)}
      \vertexSmall{(c3)}
      \vertexSmall{(c4)}
      \vertexSmall{(c8)}
    \end{tikzpicture}
    \bigg\}
    \,,
    \nn
\end{align}
where the fifth integral in this list appears with a numerator factor symbolized by $\mathcal{N}$.
Only the propagator structure (and not the vertex rules) is implied by these diagrams.
Detailed expressions, including the results of the evaluation we now outline, are given in \Cref{app:Integrals}.
We solve these integrals using the method of differential equations \cite{Henn:2013pwa}, which we will now summarize.

The only non-trivial behavior of these integrals comes from their dependence on the scattering angle $\theta$ through $x$ in \cref{eq:x}.
The differential equation for the master integrals is obtained by differentiating them with respect to this parameter and IBP reducing the result back to master integrals:
\begin{align}
  \frac{\rm d}{{\rm d}x}\vec{I}(x;\varepsilon)=M(x;\varepsilon)\vec{I}(x;\varepsilon).
\end{align}
By performing a suitable transformation $\vec{I}_c (x; \varepsilon) = T(x; \varepsilon) \vec{I}(x; \varepsilon)$, we may bring the differential equation into the canonical form
\begin{align}\label{eq:CanonicalDE}
  \frac{\rm d}{{\rm d}x}\vec{I}_{\rm c}(x;\varepsilon)=\varepsilon M_{\rm c}(x)\vec{I}_{\rm c}(x;\varepsilon),
\end{align}
solved by the path-ordered exponential
\begin{align} \label{eq:PathOrderedExponential}
  \vec{I}_{\rm c}(x;\varepsilon)=\mathcal{P}\exp\left[\varepsilon\int_{x_{0}}^{x}M_{\rm c}(x^{\prime}){\rm d}x^{\prime}\right]\vec{J}(\varepsilon).
\end{align}
Here, $\vec{J}(\varepsilon)$ are the boundary values at some boundary point $x_{0}$ where the leading behavior of the master integrals as $x\rightarrow x_{0}$, which we denote
$\vec{I}_{\rm c}^{\,(0)}$, is known;
we choose
$x_{0}=0$.
We canonicalize the differential equation using the package \texttt{CANONICA} \cite{Meyer:2017joq}.
At one and two loops, the differential equations contain only multiple polylogarithms, and the letters (i.e the canonical matrix' simple poles in $x$) are simply $\left\{-1, 0, 1\right\}$. 

Once the differential equation is canonicalized, the matrix $M_{\rm c}(x)$ admits the Laurent expansion
\begin{align}
  M_{\rm c}(x)=\frac{1}{x}M_{\rm c}^{(-1)}+\cO(x^{0}),
\end{align}
around the boundary value $x_{0}=0$. With this in hand, we may expand eq. (\ref{eq:PathOrderedExponential}) to leading order in $x$, finding\footnote{This involves the divergent integral $\int_{0}^{x}{\rm d}x^{\prime}/x^{\prime}$ which, when regularized to preserve shuffle relations, simply gives $\log(x)$ \cite{Duhr:2019tlz}.}
\begin{align}
  \vec{I}_c(x;\varepsilon)= x^{\varepsilon M_{\rm c}^{(-1)}} \vec{J}(\varepsilon) + \mathcal{O}(x),
\end{align}
or, rearranging and expanding our canonical integrals to leading order around the boundary point,
\begin{align}\label{eq:BoundaryConstants}
    \vec{J}(\varepsilon)&=x^{-\varepsilon M_{\rm c}^{(-1)}}\vec{I}_{\rm c}^{\,(0)}(x;\varepsilon).
\end{align}
Here, $\vec{J}(\epsDR)$ is independent of $x$, in contrast to the integrals' asymptotic behaviour encoded in $\vec{I}^{\,(0)}_{\rm c}(x;\varepsilon)$, which must be compensated by the matrix exponential.
Specializing to the two-loop case, the canonical master integrals pertaining to the basis \eqref{eq:2LoopMasters} are given in \cref{eq:TwoLoopCanonicalMasters}.
\Cref{eq:BoundaryConstants} then takes the explicit form
\begin{subequations}\label{eq:BoundariesTwoLoops}
\begin{align}
    \vec{J}(\varepsilon)&=
    \begin{pmatrix}
        J_{1}(\varepsilon) \\
        J_{2}(\varepsilon) \\
        J_{3}(\varepsilon) \\
        J_{4}(\varepsilon) \\
        J_{5}(\varepsilon)\\
        J_{6}(\varepsilon)
    \end{pmatrix}
    =
    \begin{pmatrix}
        I_{{\rm c},1}^{(0)} \\
        x^{4\varepsilon}I_{{\rm c},2}^{(0)} \\
        I_{{\rm c},3}^{(0)} \\
        x^{4\varepsilon}I_{{\rm c},4}^{(0)} \\
        J_{5}(\varepsilon)\\
        x^{4\varepsilon}I_{{\rm c},6}^{(0)}
    \end{pmatrix},
\end{align}
where
\begin{equation}\label{eq:J5TwoLoop}
\begin{aligned}
  J_{5}(\varepsilon)&=\left(\frac{1}{16}I_{{\rm c},1}^{(0)}-\frac{5}{8}I_{{\rm c},3}^{(0)}\right)+\left(\frac{5}{12}I_{{\rm c},2}^{(0)}+\frac{4}{3}I_{{\rm c},6}^{(0)}\right)x^{4\varepsilon} \\
  &\quad+\left(I_{{\rm c},5}^{(0)}-\frac{1}{16}I_{{\rm c},1}^{(0)}-\frac{5}{12}I_{{\rm c},2}^{(0)}+\frac{5}{8}I_{{\rm c},3}^{(0)}-\frac{4}{3}I_{{\rm c},6}^{(0)}\right)x^{-2\varepsilon},
\end{aligned}
\end{equation}
\end{subequations}
and where we have omitted the dependence of the master integrals on $x$ and $\varepsilon$.
From \cref{eq:BoundariesTwoLoops} we can immediately glean that different boundary constants are related to one another, but we must first make sense of the explicit powers of $x$ coming from the matrix exponential.
These are intimately related to the regions of the master integrals that dominate in the boundary limit.

In the $x\to0$ limit of our integrals, where the outgoing wave is parallel to the incoming one, we find only two regions for the potential (black) and active (red) gravitons with distinctive behaviors and which do not lead to scaleless integrals.
First is the \textit{forward} region, where all potential gravitons have momenta which scale as $(\ell^0_{\rm pot.},\vct{\ell}_{\rm pot.})\sim(0,x\omega)$ and all active gravitons are nearly parallel, $(\ell^0_{\rm act.},\vct{\ell}_{\rm act.})=k_1+\cO(x)=k_2+\cO(x)\sim(\omega,(1+x)\omega)$; however, as the external momenta are on-shell, the $\cO(x)$ part of active graviton propagators dominate when they appear in an integral.
Second, the \textit{general wave} region is where $(\ell^0_{\rm pot.},\vct{\ell}_{\rm pot.})\sim(0,\omega)$ and $(\ell^0_{\rm act.},\vct{\ell}_{\rm act.})\sim(\omega,\omega)$.

Despite every graviton having two possible scalings, we find that all boundary integrals up to two loops receive contributions from only two non-vanishing regions: one where all gravitons have forward scaling, and one where all gravitons have general wave scaling.
These are summarized by the conditions $\boldsymbol{\ell}_{i}\sim x\omega$ and $\boldsymbol{\ell}_{i}\gg x\omega$ for all loop momenta, respectively.
Explicitly,
\begin{align}\label{eq:BoundaryRegions}
  \lim_{x\rightarrow0}\vec{I}_{\rm c}(x;\varepsilon) 
  = 
  x^{-4\epsDR} \vec{I}_{\rm c, for.}(x;\varepsilon)
  +
  \vec{I}_{\rm c, gen.}(x;\varepsilon)
  \ ,
\end{align}
where $\vec{I}_{\rm c, for.}$ and $\vec{I}_{\rm c, gen.}$ admit Taylor expansions in $x$, and where the explicit factor of $x$ comes from the measure of integration.
In particular, a given integral in $\vec{I}_{c}^{\,(0)}$ comes from either the forward or the general wave region, and in the first case it will scale asymptotically as $x^{-4\epsDR}$.

Coming back to \cref{eq:BoundariesTwoLoops}, we thus interpret the explicit factors of $x$ as encoding the leading behavior of the canonical master integrals in the boundary limit.
Namely, since the left-hand side of this equation is independent of $x$, the leading behavior of each integral must be such that the powers of $x$ on the right-hand side cancel.
We learn that the integrals $I^{(0)}_{{\rm c}, \{2,4,6\}}$ are leading in the forward region, while the integrals $I^{(0)}_{{\rm c},\{1,3\}}$ have only a general wave region.

\Cref{eq:BoundaryRegions} also informs us that the second line of \cref{eq:J5TwoLoop} must vanish, as none of the boundary integrals possess a region which could render this line $x$-independent.
The constraint on $J_{5}$ therefore reduces to
\begin{align}
  J_{5}(\varepsilon)&=\frac{1}{16}J_{1}(\varepsilon)+\frac{5}{12}J_{2}(\varepsilon)-\frac{5}{8}J_{3}(\varepsilon)+\frac{4}{3}J_{6}(\varepsilon).
\end{align} 
The upshot of our analysis is thus two-fold.
First is the reduction in the number of boundary integrals we must compute to solve \cref{eq:CanonicalDE} for this problem: five instead of six at this loop order.
Second, we need only compute the five boundary integrals in the regions determined by \cref{eq:BoundariesTwoLoops}.

Generally speaking, in the problem at hand, the powers of $x$ which appear in \cref{eq:BoundaryConstants} are $x^{n\varepsilon}$ where $n$ are any of the eigenvalues of $M_{\rm c}^{(-1)}$.
As in the two-loop example, negative powers do not contribute to the boundary constants, while non-negative powers are related to the leading regions of the integrals in the boundary limit $x\rightarrow0$.
Up to two loops we found only two non-vanishing regions in this limit.
It will be interesting to see whether this persists at higher loop orders.

\section{Two-loop results \& BHPT matching}\label{sec:Results}

We now come to our results for the scattering amplitude up to $\cO(\epsPM^{3})$.
Our calculations pass several important checks.
Most non-trivially, our $T$-matrix element is gauge invariant.
This can be manifested by writing the amplitude in terms of field strength tensors,
\begin{align}
  \begin{aligned}
  &\hspace{2cm}
  iT
  =
  \frac1{\omega^5}
  \big(
  f_1(x) A^2
  +
  f_2(x) AB
  +
  f_3(x) B^2
  \big)
  \ ,
  \\
  &
  A\equiv F_1^\mn F^{*}_{2,\mn}
  \ ,\qquad
  B\equiv v_{\mu} F_1^\mn F^{*}_{2,\nu\rho} v^\rho
  \ ,\qquad
  F_i^\mn = \epsilon_i^{[\mu} k_i^{\nu]}
  \ ,
  \end{aligned}
\end{align}
with the asterisk representing complex conjugation.
In fact, gauge invariance enables the reconstruction of this covariant form of the $T$-matrix element from its expression in TT gauge.

Next, Weinberg's soft graviton theorem demands that the IR divergences of the $T$-matrix element exponentiate at all orders in perturbation theory \cite{Weinberg:1965nx}:
\begin{align}
    iT=\exp(-i\epsPM/\epsDR)iT_{\rm fin.},
\end{align}
Here, $T^{(n)}_{\rm fin.}$ is the unique contribution at $n$-loop order, which is $\varepsilon$-independent, and the left-hand side is truncated at $\cO(\varepsilon^{0})$.
We have confirmed that our amplitude obeys this exponentiation.
Finally, we reproduce the $iT^{(2)}$ of ref.~\cite{Bjerrum-Bohr:2025bqg}.
Our results for the $T$-matrix element are provided covariantly in the supplementary material.

Through \cref{eq:NFromT}, we find the $N$-matrix element up to $\cO(\epsPM^{3})$, presented here in the rest frame of the black hole and the kinematics in \cref{eq:Kinematics}, and decomposed in a PM expansion as
\begin{align}
  N_{h_{1},h_{2}} = \frac{e^{2i\sigma_{1}\phi}}{\omega} \sum_n  (\pi \epsPM)^n N^{(n)}_{\sigma}(x).
\end{align}
Apart from the overall phase, the helicity dependence of the amplitude is contained in $\sigma$, defined in \cref{eq:HelicityProd}.
For helicity-preserving scattering ($\sigma=1$), the amplitude is
\begin{subequations}\label{eq:NWQFTHp}
  \begin{align}
    N^{(1)}_{1}(x)&=\frac{2(x^{2}-1)^{2}}{x^{2}}\label{eq:N1WQFTHp} 
    \\
    N^{(2)}_{1}(x)&=\frac{(x-1)^{2}(15x^{2}+28x+15)}{8x(x+1)^{2}} \label{eq:N2WQFTHp}
    \\
    N^{(3)}_{1}(x)&=
    -
    \frac{134 x^4 - 19 x^2 - 91}{54\pi^{2}\left(x^2-1\right)}
    -
    2\frac{2 x^4-3 x^2+2}{3\left(x^2-1\right)^2}
    +
    \log (x)\frac{88x^{6} - 81x^{4} + 90x^{2} - 81}{18\pi^{2}\left(x^2-1\right)^2}\notag \\
    &\quad
    -
    L_{+}(x)\frac{2 x^8 + 4 x^6 - 6 x^4 + 4 x^2 + 2}{3\pi^{2}x^2 \left(x^2-1\right)^2}
    -
    L_{-}(x)\frac{15 x^6-7 x^4-7 x^2+15}{8\pi^{2}x\left(x^2-1\right)^2}
    \ .
    \label{eq:N3WQFTHp} 
  \end{align}
\end{subequations}
At two loops, the pole in the backward limit $x\rightarrow1$ is spurious.
Helicity-reversing scattering ($\sigma=-1$) takes the simpler form
\begin{subequations}\label{eq:NWQFTHr}
  \begin{align}
    N^{(1)}_{-1}(x)&=2x^{2}\label{eq:N1WQFTHr} \\
    N^{(2)}_{-1}(x)&=0\label{eq:N2WQFTHr}\\
    N^{(3)}_{-1}(x)&=-\frac{121x^{4}-450x^{2}+360}{27\pi^{2}x^2}+\log (x)\frac{4(11x^{4}-60x^{2}+60)}{9\pi^{2}x^2}\label{eq:N3WQFTHr} \\
    &\quad-L_{+}(x)\frac{2(x^{6}-12x^{4}+30x^{2}-20)}{3\pi^{2}x^4}.\nn
  \end{align}
\end{subequations}
The functions $L_{\pm}(x)$ are given in terms of multiple polylogarithms as\footnote{These are defined as $G(a_{2},a_{1};x)=\int^{x}_{\lambda_{2}}\frac{{\rm d}x_{2}}{x_{2}-a_{2}}\int^{x_{2}}_{\lambda_{1}}\frac{{\rm d}x_{1}}{x_{1}-a_{1}}$. 
Shuffle-regularizing our integrals, $\lambda_{i}=0$ if the integral is convergent there and $\lambda_{i}=1$ otherwise. Thus, \cref{eq:TwoLoopMPLs} evaluates to $L_{\pm}(x)=2{\rm Li}_{2}(-x)\pm2{\rm Li}_{2}(x)+\log(x^{2})\left[\log(1+x)\pm\log(1-x)\right]$.}
\begin{align}\label{eq:TwoLoopMPLs}
    L_{\pm}(x)&\equiv2\,G(-1,0;x)\pm2\,G(1,0;x).
\end{align}
All IR divergences in $T$ have cancelled with the iterative contributions, in addition to any scale dependence of the integrals in \Cref{app:Integrals}.

With the $N$-matrix element in hand, what's left is to verify \cref{eq:deltalWQFT}.
Up to $\cO(\epsPM^{3})$, the phase shift entering the right-hand side of this equation is \cite{Mano:1996mf,Bautista:2023sdf}
\begin{align}
{}_{-2}\delta_{\ell 2}^{P,(1)}
&= \epsPM\,
\bigg[
  \frac{3P-(2 \ell+1) \left(\ell^2+\ell-1\right)}{(\ell-1)\ell(\ell+1)(\ell+2)}
  -\psi^{(0)}(\ell-1)
  +\log(2\epsPM)
  -\frac{1}{2}
\bigg]\,,\label{eq:delta1}
\\[6pt]
{}_{-2}\delta_{\ell 2}^{P,(2)} &= \epsPM^2\pi\,
\frac{\ell(\ell+1)\big[15\ell(\ell+1)+13\big]+24}
{4\,\ell(\ell+1)(2\ell-1)(2\ell+1)(2\ell+3)}\,,\label{eq:delta2}
\\[6pt]
{}_{-2}\delta_{\ell 2}^{P,(3)} &= \frac{\epsPM^3}{12}\, 
\Bigg[
  \psi^{(2)}(\ell-1)+\psi^{(2)}(\ell+3)\nn \\
  &-\frac{432\,P}
        {\ell^3(\ell^{2}-1)^{3}(\ell+2)^{3}}
  +\frac{9\,\big(3\ell^2+3\ell+2\big)}{\ell(\ell+1)(2 \ell-1) (2 \ell+3)}
\label{eq:delta3}\\
&
    +
    \frac{3\ell(\ell+1)\big[15\ell(\ell+1)+13\big]+72}{\ell(\ell+1)
    (2\ell-1)(2\ell+1)(2\ell+3)}
      \big[\psi^{(1)}(\ell-1)+\psi^{(1)}(\ell+3)\big]
\Bigg]\nn,
\end{align}
where $\psi^{(n)}(z) = \frac{d^{\,n+1}}{dz^{\,n+1}} \ln \Gamma(z)$.
Evaluating the left-hand side of \cref{eq:deltalWQFT} (for the first several values of $\ell$ until a general-$\ell$ pattern is discernable), the one-loop -- \cref{eq:N2WQFTHp,eq:N2WQFTHr} -- and two-loop -- \cref{eq:N3WQFTHp,eq:N3WQFTHr} -- $N$-matrix elements easily agree with the NLO and NNLO phase shifts in \cref{eq:delta2,eq:delta3} respectively. 
The leading-order $N$-matrix element in the helicity-reversing case, \cref{eq:N1WQFTHr}, is also consistent with \cref{eq:delta1}.

For the leading-order helicity-preserving case, \cref{eq:TreeLevelModes} gives
\begin{align}\label{eq:TreeLevelModesExplicit}
  \frac{1}{4\pi}&\frac{N^{\varepsilon,(1)}_{\ell;h,h}}{\sqrt{\pi(2\ell+1)}} \\
  &=-\frac{ \epsPM }{\varepsilon}  +\epsPM\left[\frac{-2(2\ell+1)(\ell^{2}+\ell-1)}{(\ell -1) \ell  (\ell +1) (\ell +2)}-2 \psi ^{(0)}(\ell -1)+2 \log \left(\frac{\pi\epsPM }{Gm\bar{\mu}}\right)-\gamma_{\rm E} \right].\nn
\end{align}
As intended,  the regularized modes defined in \cref{eq:TreeLevelReg} are finite as $\varepsilon\to0$.
These modes depend on an infrared logarithm, which has an analog in \cref{eq:delta1}.\footnote{Like the logarithm in \cref{eq:TreeLevelModesExplicit}, that in the phase shift has its origin in a divergence. Specifically, the infinite sum in eq.~\eqref{eq:ThpBHPT} is divergent in the forward limit at leading order in $\epsPM$; this divergence is traditionally regularized by resumming a sub-sector of the phase-shift in $\epsPM$ \cite{Dolan:2007ut}.}
This and the remaining $\ell$-independent terms simply contribute an overall phase to the scattering amplitude, and thus, as pointed out in ref.~\cite{Ivanov:2024sds}, are unimportant in cross-section computations.
Nevertheless, these terms can be matched between both approaches by fixing the infrared scale to
\begin{align}
  \bar{\mu}&=\frac{\pi}{2Gm}e^{\frac{1-\gamma_E}{2}}.
\end{align}
In conclusion, the minimal WQFT description of a gravitating point particle produces the general relativistic scattering of a gravitational wave off of a Schwarzschild black hole up to $\cO(\epsPM^3)$.
Reading the matching backwards, the $N$-matrix element provides formulae for the infinite sums in \cref{eq:AmplitudeBHPT}.

\section{Summary \& outlook}\label{sec:conclusions}

We have leveraged the WQFT toolkit developed for the two-body problem to initiate the matching of BHPT to WQFT, using the amplitude for the scattering of a gravitational wave off of a black hole towards this end.
Focusing on the case of a spinless black hole, the amplitudes in the former formalism are derived from the Regge-Wheeler equation, and are readily available in the literature.
On the WQFT side, the starting point is the action for a spinless point particle.
With the first possible spinless non-minimal operators correcting the WQFT action entering the scattering amplitude at $\cO(\epsPM^{5})$, we expected the minimal WQFT action to produce the BHPT amplitude up to the $\epsPM^{3}$ (two-loop) order considered here, and indeed found this to be the case.

A centerpiece of our analysis has been the $\hat{N}$ operator, which furnishes the exponential representation of the $S$ matrix.
Matrix elements of this operator exhibited many advantageous features compared to $T$-matrix elements for the process under consideration.
Among these are the absence of infrared singularities and infrared scales associated with loop integration, but most crucial was an improved behavior in the forward limit.
With the exception of leading order (where the $N$- and $T$-matrix elements are equal), the $N$-matrix element is well-behaved enough as $x\rightarrow0$ to admit a decomposition in terms of four-dimensional spin-weighted spherical harmonics.\footnote{Based on the leading behavior of \cref{eq:NWQFTHp,eq:NWQFTHr} as $x\rightarrow0$, the trend indicates that higher-order contributions are even finite in this limit.}
The same cannot be said of $T$-matrix elements, which should then require a genuinely $D$-dimensional approach along the lines of ref.~\cite{Ivanov:2024sds}, or a partial-wave analysis employing spherical states as in ref.~\cite{Saketh:2023bul}.

Not only was it rather straightforward to project the $N$-matrix element onto a spherical basis, but it turned out that the modes of this projection (subject to well-understood assumptions) exponentiated.
In this very specific sense, the matrix element of the exponential \eqref{eq:SMatrixExp} can be swapped for the exponential of the matrix element.
It followed immediately that these modes were directly related to the phase shift one extracts from BHPT.

In regards to integration, aspects such as the localization of worldline energies, relatively simple differential equations, and the presence of only two regions for the boundary integrals greatly facilitated the calculation.

Having established the framework for matching WQFT to the underlying theory, we intend to pursue this matching to higher perturbative orders.
In particular, such calculations will enable the matching of Wilson coefficients of operators modifying the minimal WQFT action and encoding finite-size effects, crucial for the accurate description of binary-black hole scattering at high PM precisions. 

With these non-minimal operators entering binary scattering no sooner than the 6PM order (delayed to 7PM by the vanishing of static tides), their incorporation into spinless scattering is hindered by the current state of Feynman integration technology.\footnote{On the other hand, preliminary investigations at three and four loops hint at remarkable persisting simplicity for a single-scale problem.
It will be exciting to learn whether this process evades the computational bottlenecks of binary scattering.}
However, at a fixed PM precision, the inclusion of spin swaps loop orders for powers of the spin vector.
The impact of finite-size effects on binary scattering can thus be readily computed at lower loop but higher spin orders with present integration capabilities.
Prerequisite for such a computation is the matching of spinning WQFT (i.e. in the formulation of ref.~\cite{Haddad:2024ebn}) to dynamics described by the homogeneous Teukolsky equation; relevant work in a similar vein can be found in refs.~\cite{Dolan:2008kf,Bautista:2021wfy,Bautista:2022wjf,Bautista:2023sdf}.

Matching at higher loop orders will bring with it further scrutiny of the $N$-matrix element for this process.
In particular, it will be interesting to observe whether its improved -- compared to the $T$-matrix element -- infrared and collinear behavior persists, and additionally to understand how generic this property is.
While the $N$-matrix element is more desirable for carrying out the matching, we needed the $T$-matrix element as an intermediate quantity.
This is because a systematic procedure is lacking for the direct calculation of $N$-matrix elements; see, however, ref.~\cite{Brandhuber:2025igz} for recent progress.
A mechanism for such a direct computation would remove significant calculational redundancies, and is thus worthy of further exploration.

\section*{Acknowledgments}

We are very grateful to the authors of \cite{Ivanov:2026icp} for sharing their results for the $N$-matrix element up to two loops, and to the authors of \cite{Bjerrum-Bohr:2026fhs} for comparisons of the integrals and $T$-matrix element, and to both groups for sharing preliminary drafts of their works and coordinating the releases of their papers.
We are additionally thankful to Carl Jordan Eriksen, Benjamin Sauer, Jan Plefka, and Zihan Zhou for very helpful discussions.
This work has made use of the Black Hole Perturbation Toolkit \cite{BHPToolkit} and \texttt{FeynCalc} \cite{Mertig:1990an,Shtabovenko:2016sxi,Shtabovenko:2020gxv,Shtabovenko:2023idz}.
Integrand generation and tensor reduction was done with \texttt{FORM} \cite{Ruijl:2017dtg}.
The work of M.D. and K.H.~was funded by the European Union through the 
European Research Council under ERC Advanced Grant 101097219 (GraWFTy). The work of
Y.F.B. has been partially supported by the European Research Council under Advanced Investigator Grants ERC–AdG–885414 and  ERC–AdG–101200505.
Views and opinions expressed are however those of the authors only and do not necessarily reflect those of the European Union or European Research Council Executive Agency. Neither the European Union nor the granting authority can be held responsible for them.

\appendix

\section{WQFT framework and classical $n$-point functions}\label{app:npointfunctions}

The worldline quantum field theory approach is based on the classical $\hbar\to0$ limit of a path integral
\begin{align}\label{eq:pathintegral}
  Z[J_A]
  =
  \int D[\phi_A] \exp[
    \frac{i}{\hbar}
    \left(S
    -
    \sum_A
    \int \d^D x \phi_A(x) J_A(x)\right)
  ]
  \ ,
\end{align}
where one assumes that no variables in the action $S$ scale with $\hbar$.
Here, in order to establish the general theory, we work with arbitrary (bosonic) fields $\phi_A(x)$ labelled by a flavor index $A$.
In the context of this work, this index $A$ covers the two possibilities $h_\mn(x)$ or $z^\mu(\tau)$.\footnote{Since the graviton perturbation is a bulk field and the trajectory perturbation is localized to the worldline, the dimensionality of the integral in the source term of \cref{eq:pathintegral} technically also depends on $A$. We omit this to avoid notational clutter.}
It is well-known that in this ``naive'' $\hbar\to0$ limit the path integral localizes on tree-level contributions, which is an advantage of the WQFT framework for classical physics~\cite{Mogull:2020sak,Jakobsen:2023oow}.

Technically, in the WQFT approach, one often wishes to impose boundary condition at past infinity which formally can be achieved with the in-in framework instead of the in-out path integral given in eq.~\eqref{eq:pathintegral}.
In the classical limit, however, the in-in diagrammatics is essentially identical to the in-out one~\cite{Jakobsen:2022psy}, and we will not dwell on these details here.

The correlator of $n$ fields in the classical limit is defined by
\begin{align}
  \big\langle
    \phi_{A_1}(x_1)\dots\phi_{A_n}(x_n)
  \big\rangle
    =
  \frac{
    \int D[\phi_A] 
  \phi_{A_1}(x_1)
  \dots
  \phi_{A_n}(x_n)
  \exp(
    \frac{iS}{\hbar}
  )}
  {\int D[\phi_A]
  \exp(
    \frac{iS}{\hbar}
  )}
  \Bigg|_{\hbar\to0}
  \ .
\end{align}
Vacuum bubbles are irrelevant in the classical limit, so that the denominator trivializes and the correlator factorizes into one-point functions:
\begin{align}
    \big\langle
    \phi_{A_1}(x_1)\dots\phi_{A_n}(x_n)
  \big\rangle
  =
  \prod_i
  \big\langle
  \phi_{A_i}(x_i)\big\rangle
  \ .
\end{align}
This property allows us to show that one-point functions solve the equations of motion in this limit via the Schwinger-Dyson equations~\cite{Boulware:1968zz},
\begin{align}
  0
  =
  \bigg\langle
    \frac{\delta S[\phi_A]}{\delta \phi_A}
  \bigg\rangle
  =
  \frac{\delta S[\langle\phi_A\rangle]}{\delta \phi_A}
  \ .
\end{align}

Non-trivial $n$-point functions in the classical limit are instead derived from the logarithm of $Z[J_A]$,
\begin{align}\label{eq:logZ}
  iW[J_A]
  =
  \hbar \log(Z[J_A]) \big|_{\hbar\to0}
  \ ,
\end{align}
The upshot is that $W[J_A]$ is the generator of connected tree-level $n$-point functions through,
\begin{align}
  \big\langle
    \phi_{A_1}(x_1)\dots\phi_{A_n}(x_n)
  \big\rangle_{\rm con.}
  =
  i^n
  \frac{\delta}{\delta J_{A_1}(x_1)}
  \dots
  \frac{\delta}{\delta J_{A_n}(x_n)}
  iW[J_A]
  \bigg|_{J_A\to0}
  \ ,
\end{align}
where the $\hbar\to0$ limit was already taken in eq.~\eqref{eq:logZ}.

Finally, we note that in this classical limit, the quantum action
\begin{align}\label{eq:QuantumActionAndW}
  \Gamma
  \big[
    \phi_A[x;J_A]
  \big]
  =
  W[J_A]
  -
  \sum_B
  \int \d^D x 
  \, J_B(x) \frac{\delta W[J_A]}{\delta J_B(x)}
\end{align}
is simply the classical action, $\Gamma[\phi]=S[\phi]$.
Here, $\phi_A[x;J_A]$ is the classical field in the presence of sources $J_A$.

Through the relationship between $\Gamma=S$ and $W$ in eq.~\eqref{eq:QuantumActionAndW} one may derive equations of motion for the classical connected $n$-point functions.
We focus on the case $n=2$ which will be relevant for this work and find:
\begin{align}\label{eq:EOMForTwoPoints}
  -\delta^D(x_1-x_2)\delta_{AB}
  =
  \sum_C \int \d^D x_3
  \frac{\delta^2 i S}{\delta \phi_A(x_1)\delta\phi_C(x_3)}
    \big\langle
    \phi_C(x_3)\phi_B(x_2)
  \big\rangle_{\rm con.}
  \ .
\end{align}
The action on the right-hand-side should be evaluated on the classical fields in the absence sources, i.e. $J_A=0$.
This term with the action is a local differential operator that acts on the connected two-point function and produces a delta function.
In this sense, the connected two-point functions are Green's functions to this differential operator.
Starting from eq.~\eqref{eq:EOMForTwoPoints}, one may derive the recursive relation for the graviton two-point function in eq.~\eqref{eq:BerendsGiele}.

We note also the equation,
\begin{align}
  \frac{\delta W[J_A]}{\delta J_B(x)}
  =
  -\phi_B[x;J_A]
  \ ,
\end{align}
where, again, $\phi_B[J_A]$ is the solution of the classical equations of motion in the presence of external sources $J_A$.
Thus, in other words, this classical solution to the equations of motion $\phi_J(x)$ generates connected $n$-point functions just as well:
\begin{align}\label{eq:SourceGeneration}
  i^n
  \frac{\delta}{\delta J_{A_1}(x_1)}
  \dots
  \frac{\delta}{\delta J_{A_n}(x_n)}
  \phi_{A_0}[x;J_A]
  \Big|_{J_A\to0}
  =
  \big\langle
    \phi_{A_0}(x)
    \phi_{A_1}(x_1)
    \dots 
    \phi_{A_n}(x_n)
  \big\rangle_{\rm con.}
\end{align}

The main property of (causal) in-in theory in the classical limit is the doubling of fields.
In the Schwinger-Keldysh basis every field come in a plus and minus version: $\phi^{(\pm)}_A$.
Roughly speaking, $(+)$ may be identified with outgoing fields and $(-)$ with incoming ones.
Nonzero connected $n$-point functions then generally connect $(n-1)$ incoming fields with a single outgoing one.
In the context of this paper the connected graviton two-point function has one incoming and one outgoing leg.

\section{Integrals}\label{app:Integrals}
We list here the values of the master integrals needed at one- and two-loop order.
Because of the energy-conserving delta functions present in the integrands, the temporal components of the loop momenta are fixed to the energy of the wave.
As this is a positive quantity, the retarded and Feynman prescriptions are rendered equivalent.

At one-loop order the integral family is
\begin{align}
    I_{n_{1}n_{2}n_{3}}&=\tilde{\mu}^{2\varepsilon}\int\frac{{\rm d}^{D}\ell}{(2\pi)^{D}}\frac{\hat\delta(v\cdot(\ell-q_{1}))}{\left(\ell^{2}+ \iO\right)^{n_{1}}(\ell-q_{1})^{2n_{2}}(\ell-q_{2})^{2n_{3}}},
\end{align}
with the scale $\tilde\mu^{2}=e^{\gamma_{\rm E}}\mu^{2}/4\pi$.
Three master integrals appear at one-loop order.
These have previously been reported in ref.~\cite{Bjerrum-Bohr:2025bqg}, but we reproduce them here for completeness:
\begin{align}
    I_{100}&
    =
    \begin{tikzpicture}[baseline=7pt]
      \coordinate (cm3) at (-.5,.65) ;
      \coordinate (cm1) at (-.5,0) ;
      \coordinate (cm4) at (+1.3,.65) ;
      \coordinate (cm2) at (+1.3,0) ;
      \coordinate (c3) at (+0,.65) ;
      \coordinate (c1) at (+0,0) ;
      \coordinate (c4) at (+.8,.65) ;
      \coordinate (c2) at (+.8,0) ;
      \coordinate (c5) at (+.4,.65) ;
      \coordinate (c6) at (+.4,.5) ;
      \draw[photonTest,red] (cm3) -- (c1) ;
      \draw[photonTest,red] (c2) -- (cm4) ;
      \draw[photonTest,red] (c1) to[out=90,in=180] (c6) to[out=0,in=90] (c2) ;
      \draw[dottedLine] (cm1) -- (cm2) ;
      \vertexSmall{(c1)}
      \vertexSmall{(c2)}
    \end{tikzpicture}
    =
    -\tilde{\mu}^{2\varepsilon}(4\pi)^{\varepsilon-\frac{3}{2}}\omega^{1-2\varepsilon}e^{-i\pi\left(\frac{1}{2}-\varepsilon\right)}\Gamma\left(\varepsilon-\frac{1}{2}\right) 
    \\[2pt]
    I_{011}
    &=
    \begin{tikzpicture}[baseline=7pt]
      \draw[photonTest,red] (cm3) -- (c5) ;
      \draw[photonTest,red] (c5) -- (cm4) ;
      \draw[photonTest] (c1) -- (c5) ;
      \draw[photonTest] (c2) -- (c5) ;
      \draw[dottedLine] (cm1) -- (cm2) ;
      \vertexSmall{(c1)}
      \vertexSmall{(c2)}
      \vertexSmall{(c5)}
    \end{tikzpicture}
    =
    \frac{1}{16\omega x}+\cO(\varepsilon) \\[2pt]
    I_{111}
    &=
    \begin{tikzpicture}[baseline=7pt]
      \draw[photonTest,red] (cm3) -- (c3) ;
      \draw[photonTest,red] (c3) -- (c4) ;
      \draw[photonTest,red] (c4) -- (cm4) ;
      \draw[photonTest] (c1) -- (c3) ;
      \draw[photonTest] (c2) -- (c4) ;
      \draw[dottedLine] (cm1) -- (cm2) ;
      \vertexSmall{(c1)}
      \vertexSmall{(c2)}
      \vertexSmall{(c3)}
      \vertexSmall{(c4)}
    \end{tikzpicture}
    =
    \frac{i}{32\varepsilon\pi\omega^{3}x^{2}}\left[1-\varepsilon\log\left(\frac{4\omega^{2}x^{2}}{\mu^{2}}\right)\right]+\cO(\varepsilon).
\end{align}
Evaluating the iterations involves also the complex conjugates of these integrals.

The two-loop integral family is 
\begin{align}
    I^{\eta}_{n_{1}n_{2}n_{3}n_{4}n_{5}n_{6}n_{7}}=\tilde{\mu}^{4\varepsilon}\int\frac{{\rm d}^{D}\ell_{1}}{(2\pi)^{D}}\frac{{\rm d}^{D}\ell_{2}}{(2\pi)^{D}}\frac{\hat{\delta}(v\cdot(\ell_{1}-k_{1}))\hat{\delta}(v\cdot(\ell_{2}-k_{1}))}{D_{1}^{n_{1}}[D_{2}(\eta)]^{n_{2}}\prod_{i=3}^{7}D_{i}^{n_{i}}},
\end{align}
where the propagators are
\begin{equation}
  \begin{aligned}
    D_{1}=\ell_{1}^{2}+\iO,&\quad D_{2}(\eta)=\ell_{2}^{2}+\eta\iO, \\
    D_{i=3,4}=(\ell_{i-2}-k_{1})^{2},\quad D_{i=5,6}&=(\ell_{i-4}-k_{2})^{2},\quad D_{7}=(\ell_{1}-\ell_{2})^{2},
  \end{aligned}
\end{equation}
and $\eta=\pm1$.
Integrals in this family can be reduced to six master integrals.
Three of these can be solved exactly in $\varepsilon$:
\begin{align}
  &I_{1100000}^{\eta}
  =
    \begin{tikzpicture}[baseline=7pt]
      \coordinate (cm3) at (-.5,.65) ;
      \coordinate (cm1) at (-.5,0) ;
      \coordinate (cm4) at (+2.1,.65) ;
      \coordinate (cm2) at (+2.1,0) ;
      \coordinate (c3) at (+0,.65) ;
      \coordinate (c1) at (+0,0) ;
      \coordinate (c4) at (+.8,.65) ;
      \coordinate (c2) at (+.8,0) ;
      \coordinate (c5) at (+.4,.65) ;
      \coordinate (c6) at (+.4,.5) ;
      \coordinate (c8) at (+1.6,.65) ;
      \coordinate (c7) at (+1.6,0) ;
      \coordinate (c9) at (+1.2,.65) ;
      \coordinate (c10) at (+1.2,.5) ;
      \draw[photonTest,red] (cm3) -- (c1) ;
      \draw[photonTest,red] (c7) -- (cm4) ;
      \draw[photonTest,red] (c1) to[out=90,in=180] (c6) to[out=0,in=90] (c2) ;
      \draw[photonTest,red] (c2) to[out=90,in=180] (c10) to[out=0,in=90] (c7) ;
      \draw[dottedLine] (cm1) -- (cm2) ;
      \vertexSmall{(c1)}
      \vertexSmall{(c2)}
      \vertexSmall{(c7)}
    \end{tikzpicture}
  =\tilde{\mu}^{4\varepsilon}(4\pi)^{2\varepsilon-3}\omega^{2-4\varepsilon}e^{-(1+\eta)i\pi\left(\frac{1}{2}-\varepsilon\right)}\Gamma^{2}\left(\varepsilon-\frac{1}{2}\right) \\
  &I_{0001101}^{\eta}
  =
    \begin{tikzpicture}[baseline=7pt]
      \draw[photonTest,red] (c3) -- (c4) ;
      \draw[photonTest,red] (c8) -- (c4) ;
      \draw[photonTest] (c1) -- (c4) ;
      \draw[photonTest] (c2) -- (c4) ;
      \draw[photonTest] (c7) -- (c4) ;
      \draw[dottedLine] (cm1) -- (cm2) ;
      \vertexSmall{(c1)}
      \vertexSmall{(c2)}
      \vertexSmall{(c7)}
      \vertexSmall{(c4)}
    \end{tikzpicture}
  =-\tilde{\mu}^{4\varepsilon}\frac{(4\pi)^{2\varepsilon-3}}{(4\omega^{2}x^{2})^{2\varepsilon}}\frac{\Gamma^{3}\left(\frac{1}{2}-\varepsilon\right)\Gamma(2\varepsilon)}{\Gamma\left(\frac{3}{2}-3\varepsilon\right)} \\
  &I_{0110001}^{\eta}
  =
    \begin{tikzpicture}[baseline=7pt]
      \draw[photonTest,red] (cm3) -- (c5) ;
      \draw[photonTest,red] (c5) to[out=0,in=130] (c7) ;
      \draw[photonTest,red] (c7) -- (cm4) ;
      \draw[photonTest] (c1) -- (c5) ;
      \draw[photonTest] (c2) -- (c5) ;
      \draw[dottedLine] (cm1) -- (cm2) ;
      \vertexSmall{(c1)}
      \vertexSmall{(c2)}
      \vertexSmall{(c7)}
      \vertexSmall{(c5)}
    \end{tikzpicture}
  =-\tilde{\mu}^{4\varepsilon}
  \frac{
    (4\pi)^{2\varepsilon-3}
    }{\omega^{4\varepsilon}}
    \frac{
      \Gamma\!\left(\frac{1}{2}+\varepsilon\right)\Gamma^{2}\left(\frac{1}{2}-\varepsilon\right)\Gamma\!\left(1-4\varepsilon\right)\Gamma\!\left(2\varepsilon\right)
      }{
        \Gamma(1-2\varepsilon)\Gamma\left(\frac{3}{2}-3\varepsilon\right)
      }e^{2i\pi\eta\varepsilon}.
\end{align}
The remaining three up to the required precision are
\begin{align}
    &I_{1001101}^{\eta}
    =
    \begin{tikzpicture}[baseline=7pt]
      \draw[photonTest,red] (cm3) -- (c5) ;
      \draw[photonTest,red] (c5) -- (c8) ;
      \draw[photonTest,red] (c8) -- (cm4) ;
      \draw[photonTest] (c1) -- (c5) ;
      \draw[photonTest] (c2) -- (c5) ;
      \draw[photonTest] (c7) -- (c8) ;
      \draw[dottedLine] (cm1) -- (cm2) ;
      \vertexSmall{(c1)}
      \vertexSmall{(c2)}
      \vertexSmall{(c7)}
      \vertexSmall{(c5)}
      \vertexSmall{(c8)}
    \end{tikzpicture}
    \\
    &\hphantom{I_{1110011}^{\eta}}=
    \frac{-\frac{i\pi}{2\varepsilon}-i\pi G(-1,x)+i\pi\log\left(\frac{8\omega^{2}x^{2}}{\mu^{2}}\right)+G(1,0,x)-G(-1,0,x)}{128\pi^{2} x \omega^{2}}+\cO(\varepsilon) 
    \nn
    \\
    &I_{11-11101}^{\eta}
        =\mathcal{N}\!\times\!
    \begin{tikzpicture}[baseline=7pt]
      \draw[photonTest,red] (cm3) -- (c3) ;
      \draw[photonTest,red] (c3) -- (c4) ;
      \draw[photonTest,red] (c4) -- (c8) ;
      \draw[photonTest,red] (c8) -- (cm4) ;
      \draw[photonTest] (c1) -- (c3) ;
      \draw[photonTest] (c2) -- (c4) ;
      \draw[photonTest] (c7) -- (c8) ;
      \draw[dottedLine] (cm1) -- (cm2) ;
      \vertexSmall{(c1)}
      \vertexSmall{(c2)}
      \vertexSmall{(c7)}
      \vertexSmall{(c3)}
      \vertexSmall{(c4)}
      \vertexSmall{(c8)}
    \end{tikzpicture}
    =
    \frac{1-4\eta}{512\pi^{2}\omega^{2}\varepsilon^{2}}+\frac{\eta\left[4G(0,x)-i\pi-(\eta-4)\log\left(\frac{4\omega^{2}}{\mu^{2}}\right)\right]}{256\pi^{2}\omega^{2}\varepsilon}
    \nn \\
    &+\frac{8\eta\left[G(1,0,x)+G(-1,0,x)-G(0,x)^{2}-G(0,x)\log\left(\frac{4\omega^{2}}{\mu^{2}}\right)\right]+(1-4\eta)\log^{2}\left(\frac{4\omega^{2}}{\mu^{2}}\right)}{256\pi^{2}\omega^{2}}\nn \\
    &+i\frac{(\eta+3)G(-1,x)+(\eta-1)G(1,x)+\eta\log\left(\frac{4\omega^{2}}{\mu^{2}}\right)}{128\pi\omega^{2}}+\frac{3+8\eta}{3072\omega^{2}}+\cO(\varepsilon)
    \\[6pt]
    &I_{1110011}^{\eta}
    =
    \begin{tikzpicture}[baseline=7pt]
      \draw[photonTest,red] (cm3) -- (c3) ;
      \draw[photonTest,red] (c3) -- (c4) ;
      \draw[photonTest,red] (c4) -- (c8) ;
      \draw[photonTest,red] (c8) -- (cm4) ;
      \draw[photonTest] (c1) -- (c3) ;
      \draw[photonTest] (c2) -- (c4) ;
      \draw[photonTest] (c7) -- (c8) ;
      \draw[dottedLine] (cm1) -- (cm2) ;
      \vertexSmall{(c1)}
      \vertexSmall{(c2)}
      \vertexSmall{(c7)}
      \vertexSmall{(c3)}
      \vertexSmall{(c4)}
      \vertexSmall{(c8)}
    \end{tikzpicture}
    \\
    &\hphantom{I_{1110011}^{\eta}}=
    \frac{1}{128x^{2}\omega^{4}}\left[-\frac{1-3\eta}{8\pi^{2}\varepsilon^{2}}+\frac{1-3\eta}{4\pi^{2}\varepsilon}\log\left(\frac{4\omega^{2}x^{2}}{\mu^{2}}\right)+\frac{1-3\eta}{48}-\frac{iG(-\eta,x)}{\pi}\right. 
    \nn
    \\
    &\left.-\frac{1-3\eta}{4\pi^{2}}\log^{2}\left(\frac{4\omega^{2}x^{2}}{\mu^{2}}\right)+\frac{1-2\eta}{\pi^{2}}\left[G(1,0,x)+G(-1,0,x)\right]\right]+\cO(\varepsilon).\nn
\end{align}
The numerator of the second of these, $I_{11-11101}^{\eta}$, is symbolized by $\mathcal{N}$ in the diagrammatic representation.
Evaluating the iterations in \cref{eq:NFromT} requires also the complex conjugates of these master integrals.

Omitting powers of $\omega$ for brevity, the vector of two-loop master integrals
\begin{align}\label{eq:TwoLoopMasters}
  \vec{I}
  =
  \begin{pmatrix}
    I_{1100000},
    -I_{0001101},
    -I_{0110001},
    I_{1001101},
    I_{11-11101},
    -I_{1110011} 
  \end{pmatrix}^{T},
\end{align}
is canonicalized into 
\begin{align}\label{eq:TwoLoopCanonicalMasters}
    \vec{I}_{\rm c}
  =
  \begin{pmatrix}
    I_{1100000} \\
    -\frac{\varepsilon(6\varepsilon-1)}{5(1-2\varepsilon)^{2}}I_{0001101} \\
    -\frac{\varepsilon(6\varepsilon-1)}{5(1-2\varepsilon)^{2}}I_{0110001} \\
    \frac{\varepsilon^{2}x}{(1-2\varepsilon)^{2}}I_{1001101} \\
    \frac{\varepsilon^{2}}{(1-2\varepsilon)^{2}}I_{11-11101} \\
    -\frac{\varepsilon^{2}x^{2}}{(1-2\varepsilon)^{2}}I_{1110011} 
  \end{pmatrix}
\end{align}
In particular, the transformation matrix that renders the masters into canonical form is diagonal.

\section{Spin-weighted spherical harmonics}\label{app:SWSH}

We collect here pertinent conventions and relations relating to spin-weighted spherical harmonics.

Spin-weighted spherical harmonics furnish a basis for functions defined on the sphere.
As such, they are orthonormal,
\begin{align}\label{eq:SWSHOrtho}
  \int{\rm d}\Omega\, {}_{s'}\bar{Y}_{l' m'}(\theta, \phi) \, {}_sY_{l m}(\theta, \phi) = \delta_{s s'} \delta_{l l'} \, \delta_{m m'},
\end{align}
where $\int{\rm d}\Omega=\int_{0}^{2\pi}\int_{0}^{\pi}\sin\theta\,{\rm d}\theta\,{\rm d}\phi$.
Any function $f$ of $\theta\in[0,\pi],\,\phi\in[0,2\pi)$ can be written as
\begin{align}
  f(\theta,\phi)=\sum_{l=|s|}^{\infty}\sum_{m=-l}^{l}{}_{s}Y_{lm}(\theta,\phi)f_{lm}.
\end{align}
Through the orthonormality property of the spherical harmonics, the modes of the function are determined through
\begin{align}
  f_{lm}=\int{\rm d}\Omega\,{}_{s}\bar{Y}_{lm}(\theta,\phi)f(\theta,\phi).
\end{align}
The conjugate harmonic in this equation can be replaced by an unconjugated one through
\begin{align}\label{eq:SWSHConj}
  {}_{s}\bar{Y}_{lm}(\theta,\phi)=(-1)^{s+m}{}_{-s}Y_{l(-m)}(\theta,\phi).
\end{align}
The dependence of the harmonics on the azimuthal angle is simply a phase,
\begin{align}
  {}_{s}Y_{lm}(\theta,\phi)=e^{im\phi}{}_{s}Y_{lm}(\theta,0).
\end{align}
In the body of the paper, whenever the second argument is omitted, it is assumed to be 0, ${}_{s}Y_{lm}(\theta)\equiv{}_{s}Y_{lm}(\theta,0)$.

The reflection properties of the harmonics are summarized by
\begin{align}\label{eq:SWSHRefl}
  {}_{s}Y_{lm}(\theta,\phi)=(-1)^{l}\,{}_{-s}Y_{lm}(\pi-\theta,\phi+\pi).
\end{align}
Intuitively, if a unit vector $\boldsymbol{n}$ is given by the coordinates $(\theta,\phi)$, then $-\boldsymbol{n}$ has the coordinates $(\pi-\theta,\phi+\pi)$.

A final essential property is the addition theorem for spin-weighted spherical harmonics \cite{Monteverdi:2024xyp}:
\begin{align}
    (-1)^{s}\sum_{m=-l}^{l}{}_{s_{1}}\bar{Y}_{lm}(\theta_{1},\phi_{1})\,{}_{s_{2}}Y_{lm}(\theta_{2},\phi_{2})&=\sqrt{\frac{2l+1}{4\pi}}e^{-is_{2}\alpha}{}_{s_{2}}Y_{l(-s_{1})}(\beta,\gamma),
\end{align}
valid for $l\geq\max(|s|,|s^{\prime}|)$ and where $\alpha,\beta,\gamma$ are the Euler angles, which appeared in the main text but which we reproduce here for convenience:
\begin{align}
  \cot\alpha&=\cos\theta_{2}\cot(\phi_{2}-\phi_{1})-\cot\theta_{1}\sin\theta_{2}\csc(\phi_{2}-\phi_{1})\label{eq:appEulera} \\
  \cos\beta&=\cos\theta_{1}\cos\theta_{2}+\sin\theta_{1}\sin\theta_{2}\cos(\phi_{2}-\phi_{1})\label{eq:appEulerb} \\
  \cot\gamma&=\cos\theta_{1}\cot(\phi_{2}-\phi_{1})-\cot\theta_{2}\sin\theta_{1}\csc(\phi_{2}-\phi_{1}).\label{eq:appEulerg}
\end{align}
Multiplying both sides by ${}_{s_{1}}Y_{l^{\prime}m^{\prime}}(\theta_{1},\phi_{1})$ with $|m^{\prime}|\leq\min(\ell,\ell^{\prime})$ and integrating over the solid angle ${\rm d}\Omega_{1}$ of $\theta_{1},\phi_{1}$:
\begin{align}\label{eq:SWSHAdd}
    \int{\rm d}\Omega_{1}\,e^{-is_{2}\alpha}{}_{s_{2}}Y_{l(-s_{1})}(\beta,\gamma)\,{}_{s_{1}}Y_{l^{\prime}m^{\prime}}(\theta_{1},\phi_{1})
    &=(-1)^{s_{2}}\sqrt{\frac{4\pi}{2l+1}}\,{}_{s_{2}}Y_{lm^{\prime}}(\theta_{2},\phi_{2})\delta_{ll^{\prime}}.
\end{align}
The Euler angles appearing here have intuitive physical interpretations not apparent from \cref{eq:appEulera,eq:appEulerb,eq:appEulerg}.
Understanding them is central to understanding the general form in \cref{eq:AmplitudeEuler}.
Let us explore them further now.

\subsection{Euler angles}\label{app:EulerAngles}

Consider two unit vectors $\boldsymbol{n}_{1}=\boldsymbol{n}_{1}(\theta_{1},\phi_{1})$ and $\boldsymbol{n}_{2}=\boldsymbol{n}_{2}(\theta_{2},\phi_{2})$.
The angle between these vectors is easily shown to be $\boldsymbol{n}_{1}\cdot\boldsymbol{n}_{2}=\cos\beta$.

Now, the vector $\boldsymbol{n}_{i}$ induces a coordinate system $(\boldsymbol{e}_{i}^{x},\boldsymbol{e}_{i}^{y},\boldsymbol{n}_{i})$, with
\begin{align}
    \boldsymbol{e}_{1}^{x}&=\frac{\partial}{\partial\theta_{1}}\boldsymbol{n}_{1},\quad \boldsymbol{e}_{1}^{y}=\frac{1}{\sin\theta_{1}}\frac{\partial}{\partial\phi_{1}}\boldsymbol{n}_{1}, \\
    \boldsymbol{e}_{2}^{x}&=\frac{\partial}{\partial\theta_{2}}\boldsymbol{n}_{2},\quad \boldsymbol{e}_{2}^{y}=\frac{1}{\sin\theta_{2}}\frac{\partial}{\partial\phi_{2}}\boldsymbol{n}_{2}.
\end{align}
With this construction, the remaining Euler angles represent the azimuthal angle of $\boldsymbol{n}_{1}$ ($\boldsymbol{n}_{2}$) in the frame induced by $\boldsymbol{n}_{2}$ ($\boldsymbol{n}_{1}$).
Specifically
\begin{align}
  \boldsymbol{n}_{1}\cdot\boldsymbol{e}_{2}^{x}=\cos\alpha,&\quad \boldsymbol{n}_{1}\cdot\boldsymbol{e}_{2}^{y}=-\sin\alpha, \\
  \boldsymbol{n}_{2}\cdot\boldsymbol{e}_{1}^{x}=\cos\gamma,&\quad \boldsymbol{n}_{2}\cdot\boldsymbol{e}_{1}^{y}=\sin\gamma.
\end{align}
Changing to the circular basis,
\begin{align}
  \boldsymbol{\varepsilon}^{\sigma}=\frac{\boldsymbol{e}^{x}+i\sigma\boldsymbol{e}^{y}}{\sqrt{2}},
\end{align}
for $\sigma=\pm1$, one finds that these two angles are nothing but little group phases in an amplitude consisting of these polarizations and the $\boldsymbol{n}_{i}$:
\begin{align}
  \boldsymbol{n}_{1}\cdot\boldsymbol{\varepsilon}_{2}^{\sigma}=\frac{1}{\sqrt{2}}e^{-i\sigma\alpha},\quad \boldsymbol{n}_{2}\cdot\boldsymbol{\varepsilon}_{1}^{\sigma}=\frac{1}{\sqrt{2}}e^{i\sigma\gamma},\quad \boldsymbol{\varepsilon}_{1}^{\sigma_{1}}\cdot\boldsymbol{\varepsilon}_{2}^{\sigma_{2}}=\frac{1}{2}\frac{e^{i(\sigma_{1}\gamma-\sigma_{2}\alpha)}}{\cos\beta+\sigma_{1}\sigma_{2}}.
\end{align}
The only non-trivial angular dependence of such an amplitude is therefore on $\beta$.
Note that the poles in $\beta$ of the last dot product are spurious, and disappear when this product is expressed in terms of the original angles.

\bibliographystyle{JHEP}
\bibliography{TwoLoopCompton.bib}

\end{document}